\documentclass[sigconf, nonacm]{acmart}

\usepackage{listings}
\usepackage{xcolor}
\definecolor{commentgray}{HTML}{696969}
\definecolor{codegray}{rgb}{0.5,0.5,0.5}
\definecolor{backcolor}{rgb}{0.95,0.95,0.92}
\definecolor{mygreen}{HTML}{009B55}
\lstdefinestyle{mystyle}{
basicstyle=\ttfamily\scriptsize,
   backgroundcolor=\color{backcolor},
   language=Python,
   commentstyle=\color{commentgray},
   keywordstyle=\color{blue},
   numbers=none,
   breakatwhitespace=false,        
   breaklines=true,                
   captionpos=b,                    
   keepspaces=true,                
   numbers=left,                    
   numbersep=6pt,                  
   showspaces=false,                
   showstringspaces=false,
   showtabs=false,                  
   tabsize=2
}
\usepackage{cprotect}

\usepackage{graphicx}
\usepackage{multirow}

\usepackage{xspace}
\usepackage{times}
\usepackage{array}
\usepackage[normalem]{ulem}

\newcommand\vldbdoi{10.14778/3598581.3598600}

\newcommand\vldbavailabilityurl{https://github.com/ferdiko/vetl}

\newcommand{\vetl}{\emph{V-ETL}\xspace}
\newcommand{\Ours}{\emph{Skyscraper}\xspace}
\newcommand{\accbar}{q\widehat{ua}l}
\newcommand{\para}[1]{\paragraph{\textbf{#1}}}

\newcommand{\blfootnote}[1]{%
  \begingroup
  \renewcommand\thefootnote{}\footnote{#1}%
  \addtocounter{footnote}{-1}%
  \endgroup
}

\newcommand{\thickhline}{%
	\noalign {\hrule height 1pt}
}

\begin{document}
\title{Extract-Transform-Load for Video Streams}

\author{Ferdi Kossmann$^1$  Ziniu Wu$^1$ Eugenie Lai$^1$ Nesime Tatbul$^{1,2}$  Lei Cao$^{1,3}$ Tim Kraska$^{1,4}$ Sam Madden$^1$}
\email{{ferdik, ziniuw, eylai, tatbul, lcao, kraska, madden}@csail.mit.edu}
\affiliation{%
 \institution{$^1$MIT CSAIL, $^2$Intel Labs, $^3$University of Arizona, $^4$AWS}
}

\def \authors{Ferdi Kossmann, Ziniu Wu, Eugenie Lai, Nesime Tatbul, Lei Cao, Tim Kraska, Sam Madden}

\begin{abstract}
Social media, self-driving cars, and traffic cameras produce video streams at large scales and cheap cost. However, storing and querying video at such scales is prohibitively expensive. We propose to treat large-scale video analytics as a data warehousing problem: Video is a format that is easy to produce but needs to be transformed into an application-specific format that is easy to query. Analogously, we define the problem of Video Extract-Transform-Load (\vetl). \vetl systems need to reduce the cost of running a user-defined \vetl job while also giving throughput guarantees to keep up with the rate at which data is produced. We find that no current system sufficiently fulfills both needs and therefore propose \Ours, a system tailored to \vetl. \Ours can execute arbitrary video ingestion pipelines and adaptively tunes them to reduce cost at minimal or no quality degradation, e.g., by adjusting sampling rates and resolutions to the ingested content. \Ours can hereby be provisioned with cheap on-premises compute and uses a combination of buffering and cloud bursting to deal with peaks in workload caused by expensive processing configurations. In our experiments, we find that \Ours significantly reduces the cost of \vetl ingestion compared to adaptions of current SOTA systems, while at the same time giving robustness guarantees that these systems are lacking.
\end{abstract}

\maketitle
\pagestyle{plain}

\begingroup

\ifdefempty{\vldbavailabilityurl}{}{
\vspace{-0.12cm}
\begingroup\small\noindent\raggedright\textbf{PVLDB Artifact Availability:}
The source code, data, and/or other artifacts have been made available at \url{\vldbavailabilityurl}.
\endgroup
\vspace{-0.05cm}
}

\section{Introduction}
\label{sec:introduction}

Every day, millions of video streams are produced by smartphones, TV stations, self-driving cars, dashcams, and CCTV cameras deployed in cities and office buildings. These video streams can offer great insights and enormous value in fields such as city planning, marketing, advertisement, smart retail, or autonomous driving. For example, city planners around Vancouver are currently facing the challenge of deciding where to place electric vehicle (EV) chargers. For that, they want to obtain data that tells them which points in the city are most commonly traversed by EVs. Most cities like Vancouver already installed hundreds to thousands of traffic cameras, which could be used to obtain such EV counts. 

The naive way of counting how many EVs pass by each camera is to  store the video from all cameras and then run an object detection algorithm\footnote{In Canada (as in many other countries), EVs are especially easy to distinguish from other cars since they have green license plates.} on the recorded video at query time. However, this approach has major disadvantages. 
First, storing the video requires outrageously large storage volumes. For example, one thousand traffic cameras roughly produce 230 TB of data every month.\footnote{One traffic camera feed in our experiments produces 7.8GB of data per day.} 
Storing one month's data on Amazon S3 would therefore cost \$60,000 per year. 
Second, querying for trends or averages usually requires analyzing months to years of data, which leads to large query latencies. Even on modern GPUs,  state-of-the-art computer vision (CV) models can only process a few frames per second. For example, processing one year of video with the YOLO object detector~\cite{yolo} takes six months on an AWS p3.2xlarge instance (with an NVIDIA Tesla V100 GPU).
Third, naively applying CV techniques at such scales is prohibitively expensive for many applications. For example, naively running the YOLO object detector~\cite{yolo} to analyze a month of traffic data from 100 cameras costs \$110,000 on AWS\footnote{E.g., using 50 p3.2xlarge instances, each of which currently costs 3.06 USD/h.}. 

To address the limitations of the naive approach, we propose to manage live video streams like in a data warehouse. Video is a format that is easy to produce but hard to query. A \textit{video warehouse} allows for efficient querying by converting incoming video into an intermediate format that is easy to query. This intermediate format is application-specific and contains the extracted entities of interest.
In the EV example, it would contain car counts and types. 
Analogous to traditional data warehouses, we refer to the process of preparing the data for querying as Video Extract-Transform-Load (\vetl).
Video is \textit{extracted} from the cameras, \textit{transformed} into the intermediate format using CV, and \textit{loaded} into a query engine like a relational database system. This lets the user issue queries in SQL against tables with the extracted entities (e.g., obtaining the EV counts is a simple count query on a \verb|Detections| table, where the detected car is an EV, grouped by the camera id). 

Video warehouses eliminate the storage problem since users may throw video away after extracting all entities of interest during ingestion.
They also solve the query latency issue, since users can issue queries against the intermediate format and no expensive CV algorithm needs to be run at query time.
However, video warehouses do not magically solve the cost problem, as the video still needs to be processed during the \vetl Transform step. Furthermore, video processing must happen at the rate at which the video is produced in order to achieve continuous ingestion.

To address the challenges imposed by \vetl, we built \Ours which allows for cheap video ingestion while also adhering to throughput requirements. \Ours's goal is to make the \vetl \emph{transform} step more practical. It allows users to provision hardware resources according to their monetary budget and optimizes the quality of the extracted video entities on the given resources. 

Depending on the provisioned hardware, \Ours reduces the work imposed by the \vetl job while degrading the result quality as little as possible. \Ours does this by dynamically configuring knobs that are inherent to CV workloads. Examples of such knobs include the frame rate or the image resolution at which the video is processed, as well as further, application-specific knobs. Each of these knob represents a trade-off between work and result quality: Expensive knob configurations can reliably deliver good results, even for difficult inputs (e.g., many object occlusions); cheap configurations, on the other hand, only deliver good results on easy inputs (e.g. few occlusions, good lighting conditions etc.) but are prone to mistakes on difficult inputs. The content of real-world video streams is highly variable with frequent changes in how difficult it is to analyze the content (i.e., changes every few 10s of seconds). \Ours saves work by using expensive knob configurations on difficult video segments and cheap configurations otherwise.

Since \Ours needs to process data on constrained hardware at a required throughput, \Ours must configure the knobs not only based on the video content but also on the available hardware resources. Industrial deployments for live video processing are typically provisioned with three types of resources~\cite{videoedge}: a local compute cluster with a high-bandwidth connection to the video source, a video buffer, and cloud resources that may be used to rent on-demand cloud compute (to limit cloud costs, users typically want to set a cloud budget.) \Ours leverages all three of these resource types: \Ours itself runs on the local cluster and uses it to process video. To keep costs low, the local cluster is typically not provisioned to process the most expensive knob configurations in real-time. 
When it falls behind, \Ours sets video video in the buffer and, as the buffer starts to fill, offloads work to on-demand cloud workers.  

\Ours must avoid prematurely using up buffer space and cloud credits in order to not run out of them when expensive knob configurations would have the greatest impact. \Ours therefore forecasts the workload and rations compute resources with regard to future demand. 
To still be robust to unavoidable inaccuracies in the forecast, we propose to combine a predictive planning component with a reactive execution component, which lets \Ours make tuning decisions while considering both, the future demand and the content that is actually streamed in the moment. 


Despite the need for predictive knob tuning, \Ours's knob tuning decisions must impose a low overhead --- this is especially important in low-budget regimes, where large decision overheads would consume a significant portion of the compute resources. 
While prior content-adaptive knob tuners run additional CV operators to make tuning decisions~\cite{chameleon, zeus}, \Ours adapts to the video content only based on a user-defined quality metric (e.g., certainties commonly reported by CV models) that are extracted anyways when running the \vetl job. 
This allows \Ours to make tuning decisions in under 0.5 ms on a single CPU core. 


\begin{figure}[t]
    \centering
    \includegraphics[width=0.46\textwidth]{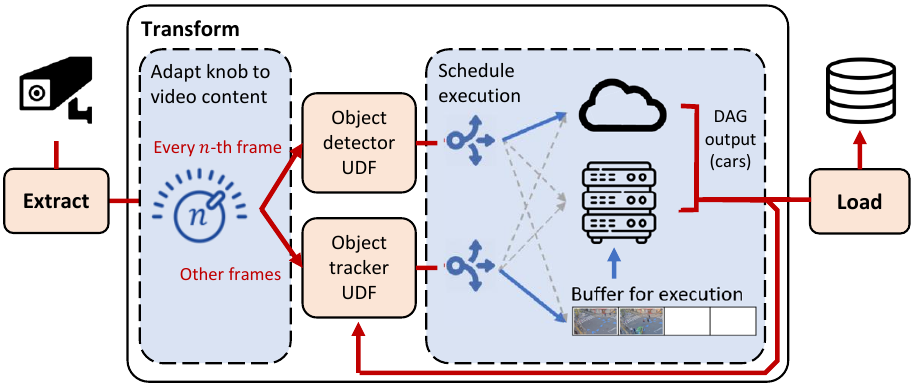}    
    \vspace{-0.5em}
    \caption{\Ours optimizing the expensive \vetl Transform step of the EV counting example job. The blue components are provided by \Ours, the red ones are provided by the user.}
    \vspace{-2em}
    \label{fig:introdag}
\end{figure}

Figure~\ref{fig:introdag} shows an overview of how \Ours processes the EV example workload. 
The user specifies user-defined functions (UDFs) that transform the video into the application-specific query format. In Figure~\ref{fig:introdag}, the user only defines two UDFs. The object detector UDF is responsible for detecting new cars, while the object tracker UDF is responsible for tracking cars as they move across the frame to avoid double counting them. Finally, the user registers the workload's tunable knobs. In the simple example, the user only defines a single knob that controls how frequently the object detector should be run. \Ours optimizes the costly Transform step while the user code performs the Extract and Load steps.

\para{Prior work} While \Ours is the first system to specifically address the challenge of \vetl, there are several lines of work that are relevant to \Ours. We briefly highlight two of them here and refer to Section~\ref{sec:rel} for a more detailed discussion on related work. Table~\ref{table:rel} shows the differences between \Ours and these works.

First, there is prior work on content-adaptive knob tuning, such as Chameleon~\cite{chameleon} and Zeus~\cite{zeus}. These systems are designed to reduce the average processing time per frame while assuming that the provisioned hardware can always ingest video in real-time (even during peak workload). However, when ingesting video on cheaper machines that are not peak-provisioned, prior systems do not provide throughput guarantees and are therefore impractical for \vetl. Adapting these systems to fulfill throughput requirements on cheap hardware is challenging, since they are agnostic to lag and the hardware resources they run on. We discuss this further in Section~\ref{sec:rel}.


Second, there is prior work on systems that use knob tuning to adapt to the current query load. VideoStorm~\cite{videostorm} and VideoEdge~\cite{videoedge} are designed for scenarios where users run a dynamic set of queries over video streams, which causes dynamic changes to the type and number of queries running. At times when many queries are running concurrently, not all queries may be able to run at maximum quality and in real time. VideoStorm and VideoEdge tune the queries' knobs such that the queries fulfill their quality and latency goals as well as possible. However, VideoStorm and VideoEdge only adapt to the query load (i.e., the queries present in the system) and are agnostic to the streamed content.
This brings no benefit in scenarios where the query load is static. While we envision most \vetl applications to ingest video using a static set of processing jobs, VideoStorm might still be used if users dynamically redefine how to ingest video. 


\begin{table}[h]
\setlength\tabcolsep{2pt}  
    \begin{tabular}{|m{1.8cm} | m{2.78cm} | m{2.47cm} | }

    \hline

    & \raggedright \small \textbf{Adapt to video content} & \raggedright \small \textbf{Adapt to query load} \tabularnewline \hline

    \raggedright \small \textbf{Throughput guarantees} & \centering \small \Ours & \centering \small VideoStorm, VideoEdge 
    \tabularnewline
    \hline

    \raggedright \small \textbf{No throughput guarantees} & \small \centering Chameleon, Zeus & \tabularnewline
    \hline
    
    \end{tabular}
    
    \vspace{0.5em}
    \caption{\Ours compared to other video knob tuning systems}
    \label{table:rel} 
    \vspace{-2em}
\end{table}

In summary, our contributions are as follows:

$\bullet$ We  define the problem of Video Extract-Transform-Load (\vetl) and identify its importance.  

$\bullet$ To make \vetl more practical, we propose \Ours, the first content-adaptive knob tuning system with throughput guarantees. \Ours lets users provision compute resources according to their budget and optimizes the result quality on the given resources.

$\bullet$ To effectively ration compute resources over time, we propose a combination of predictive planning and reactive execution.

$\bullet$ We propose a tuning method that only relies on a user-defined quality metric which is extracted anyways when running the \vetl job. We find that this method allows for negligible tuning overheads.

$\bullet$ We conduct experiments on several real-world and synthetic workloads and find that \Ours can achieve cost reductions up to $8.7\times$ over baselines on various workloads.

\section{Problem definition and System overview}
\label{sec: overview}


\subsection{Problem definition}
\label{subsec: def}

Video Extract-Transform-Load (\vetl) refers to extracting entities of interest from a video stream by processing it according to a user-defined specification and adhering to two constraints. First, \vetl systems must process video at the rate at which it arrives. A \vetl system may lag behind on processing but may only do so by a constant amount. In practice, this means that \vetl systems may use a fixed-size storage medium (i.e., buffer) to set video aside for later processing. Equation~\ref{equ: vetl} states that the size of the buffered frames may not exceed the size of the buffer.

\vspace{-1em}
\begin{align}
\label{equ: vetl}
    out(t) \subseteq in(t) \,\,\land\,\, \sum_{F \in \, in(t) \setminus \, out(t)} size(F) \leq B \hspace{3em} \forall t
\end{align}

\noindent where $t$ is a timestamp, $in(t)$ is the set of frames that the video source has produced at time $t$, $out(t)$ is the set of frames that the \vetl system has processed at time $t$, $size(F)$ is the size of frame $F$ in bytes and $B$ is the buffer size in bytes.

Second, \vetl systems must process video at a budget that is defined by the user. This budget is provided as a dollar cost $budget_T$ that may be spent over a given time interval $T$. The processing cost over interval $T$ encompasses all costs including average wear of hardware, cloud costs, etc. 
The summed cost of processing all frames in $T$ must be below $budget_T$: $\sum_{F\in T} cost(F) \leq budget_T$.


The combination of processing video at a required throughput while being constrained on computing resources makes for exciting optimization problems. \Ours aims to maximize the overall result {\it quality} by tuning workload-specific \textit{knobs} that are inherent to computer vision workloads (e.g., the frame rate or image resolution). In \Ours, the quality is user-defined and is measured and returned by the user code --- this lets \Ours generalize to different workloads with different notions of quality.

Users may further register arbitrary knobs together with a corresponding \textit{knob domain}. The knob domain is a user-defined set of values that the knob may take (e.g. the knob domain for the frame rate knob might be \{15 FPS, 30 FPS\}). \Ours dynamically configures registered knobs based on the streamed video content and maximizes the quality (e.g. accuracy) of the extracted entities while adhering to the \vetl requirements. 

Formally, a knob configuration $k$ refers to an instantiation of each knob to a value in its domain. Some knob configurations induce more work than others. Similarly, some produce more qualitative results than others. However, the result quality of a knob configuration depends not only on the configuration but also on the video content. While a high image resolution may reliably produce good results, it may not always be needed as some content can also be accurately processed at a lower resolution. Let a \textit{video segment} denote a sequence of successive frames of the video (e.g., 2 seconds of video). We denote the quality that a knob configuration $k$ achieves on a video segment $s$ as $qual(k,s)$. 
The optimization goal of \Ours is to maximize the overall quality $qual(v)$ of entities extracted from video $v$, which is given by $qual(v) = \sum_{s\in v} qual(k_s, s)$where $k_s$ is the configuration used to process segment $s$.



\subsection{System overview}
\label{subsec: overview}

The following subsection gives a high-level overview of \Ours. Section~\ref{sec: offline} and Section~\ref{sec: online} then provide a more detailed discussion of \Ours's design. We focus on how \Ours ingests a single video stream and show in Appendix~\ref{app:multistream}, how this approach can easily be generalized to multiple streams.



\begin{figure*}[t]
\vspace{-1em}
  \includegraphics[width=0.97\textwidth]{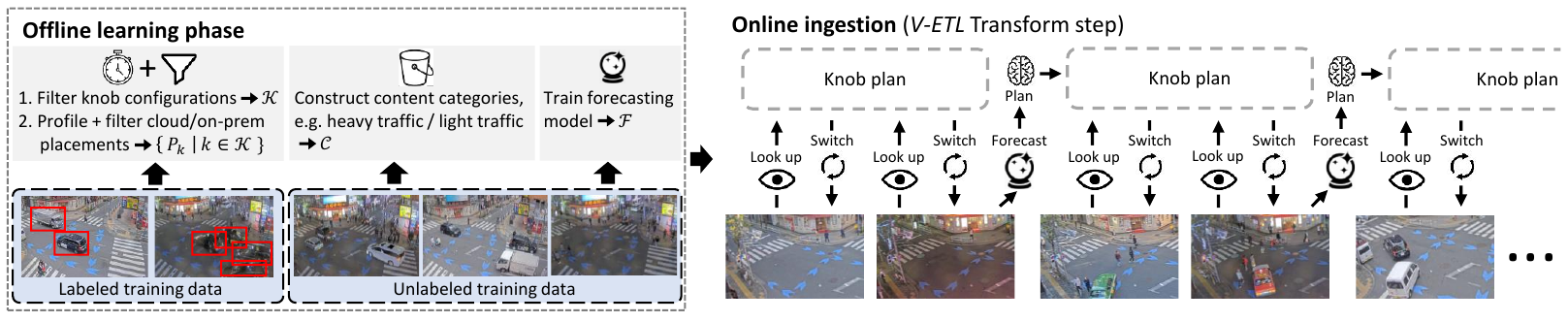}
  \vspace{-1em}
  \caption{Overview over all processing steps of \Ours.}
  \label{fig:overview}
  \vspace{-1.5em}
\end{figure*}

\vspace{0.4em}

\textbf{\textit{Design challenges.}}
To explain why \Ours works the way it does, we present a simplistic, idealized approach to content-adaptive knob tuning with throughput guarantees, and show where this approach fails in practice. We then present the ideas that \Ours uses to overcome the issues of the idealized approach. 

For now, we do not consider buffering or the scheduling of computation between on-premise resources and the cloud. Instead, we simply consider a computation budget $budget_T$ on the number of arithmetic operations that we may use to ingest video produced during time period $T$. We are further given a small set of knob configurations $\mathcal{K}$ which allows us to process different segments of the video at different costs and qualities (see Section~\ref{subsec: def}).

We observe that the knob tuning system must speculate about the future content of the video in order to effectively ration $budget_T$ over time. Otherwise, the system can not assess whether it is sensible to process content with an expensive knob configuration now or to save the budget for the future when expensive knob configurations might have a larger impact. Furthermore, we find that the effectiveness of different knob configurations often changes within seconds --- a content-adaptive knob tuning system should therefore reassess which configuration to use every couple of seconds.

Now, suppose we have a forecasting function that can perfectly predict what quality each knob configuration achieves at any given time in the future. In this idealized world, we can easily build a system that achieves optimal performance: Our optimal system would slice time interval $T$ into segments $t_i$ of equal length, where each segment $t_i$ is a few seconds long. The system then forecasts the quality that each knob configuration achieves on each segment $t_i \in T$. Finally, given the forecasted qualities, optimizing the assignment of knob configurations to segments is an instance of the 0-1 knapsack problem, where the overall quality must be maximized under the given budget $budget_T$ (more details in Appendix~\ref{app:design}).

Unfortunately, we find that achieving good accuracy on this forecasting task is infeasible in the real world. 
To forecast the knob configurations' qualities for each $t_i \in T$, our forecasting function needs to predict what happens at each second in the video, hours into the future. This is impossible since the precise timing of events is subject to substantial randomness. For example, it is impossible to predict the exact moment in which a large group of pedestrians will pass by a camera, hours into the future. To make our system work in the real world, we need to design a more practical forecasting task. 

We rely on two insights that guide the design of this new forecasting task. First, we observe that there are a few types of video content that characterize any of the videos seen throughout the live stream (e.g., rush hour traffic, normal traffic, low traffic). For the content of the same kind, each knob configuration produces results of similar quality. For example, for content with many occlusions (e.g., rush hour), knob configurations that cannot handle occlusions will always produce low-quality results. Second, we observe that, while it is impossible to predict \textit{when} certain content appears, it is possible to predict \textit{how often} it appears, {\it assuming the future video is distributed roughly as a recent historical video has been}. For example, while it is impossible to predict the \textit{precise moments} (i.e. the $t_i$'s) at which groups of pedestrian pass by the camera, it is possible to estimate \textit{how often} groups of pedestrians pass by the camera.

We can now design a forecasting task where accurate predictions are feasible in practice. Based on the first insight (content falls into a few categories), we use a simple clustering mechanism to compute \textit{content categories} such that all streamed content falls into one of these categories. We construct them such that all knob configurations achieve a similar quality on the content of the same category (more details in Section~\ref{sec: offline}). 
Then, based on the second insight (content distribution is predictable), we simply predict how often each content category appears within a time interval $T$. For example, if our forecasting model thinks that 10\% of the video in $T$ shows rush-hour traffic, it would forecast 10\% for the rush-hour category. 
In practice, we can achieve high forecasting accuracy on real-world workloads.

Finally, we need to re-think how to use the forecast for knob tuning. Since we no longer forecast the qualities of individual segments $t_i$, we cannot assign knob configurations the same way as in our idealized system. Instead, we can only assign knob configurations to content categories. Knowing how often each content category appears allows us precisely determine the overall cost of using a knob configuration to process the content of that category. In Section~\ref{sec: online}, we describe how this allows us to find the optimal assignment of knob configurations to content categories under a given budget and for a given forecast. Given this assignment, we then need to reactively determine what category the current content belongs to. Once we determine the category, we can simply look up and use the knob configuration we assigned to this category. Section~\ref{sec:knob-switch} describes a simple method for determining the current content category, which runs fast and determines the correct category with high accuracy.

In summary, we took a simplistic, idealized system and made it practical by re-designing the forecasting task. We then built an efficient system around it that can leverage this forecast for predictive knob tuning. \Ours takes these ideas and implements them for real hardware provisionings.

\vspace{0.5em}
\textbf{\textit{Skyscraper walk-through.}}
Given these challenges imposed by content-adaptive knob tuning with throughput guarantees, we now give an overview on how \Ours uses these ideas when provisioned with real hardware (i.e., with a local compute cluster, video buffer and cloud credits). 
\Ours is split into an \textit{offline learning phase} and an \textit{online ingestion phase} as shown in Figure~\ref{fig:overview}. Section~\ref{sec: offline} gives a detailed description of the offline phase and Section~\ref{sec: online} gives a detailed description of the online phase. 

The offline phase is used to pre-compute invariant properties of the \vetl workload, which allow online ingestion at negligible overheads. To compute these properties, the user provides \Ours with a small set (e.g. 5 minutes) of labeled data and a larger set (e.g. two weeks) of unlabeled data from the ingested video source. \Ours uses this data to prepare online ingestion in three steps. 

First, \Ours profiles different knob configurations on the provisioned on-premise hardware and cloud hardware. Each knob configuration corresponds to a directed acyclic graph (DAG) of UDFs.  
\Ours profiles the cloud cost and runtime of different UDF placements --- executing some UDFs on the cloud may reduce the execution time (due to added parallelism) but increases the cloud cost. \Ours filters out placements that do not lie on the cost-runtime Pareto frontier. Similarly, \Ours filters out knob configurations that do not lie on the runtime-quality Pareto frontier. 
Appendix~\ref{app:api} disusses how \Ours handles data movement between UDFs and between on premises and the cloud.

Second, \Ours uses the unlabeled data to construct the content categories as discussed under \textit{Design challenges}. The content categories are constructed solely based on a quality metric that is measured and returned by the user code (e.g. certainty or errors commonly reported by CV models). By construction, the content categories discriminate between any content characteristic that affects the quality of at least one knob configuration. Constructing the content categories solely based on a user-defined quality metric lets \Ours generalize across workloads since \Ours doesn't need to understand the precise workings of the UDFs and how their performance is affected by pixel-level changes. Furthermore, dealing with low-dimensional quality vectors (e.g., 5-dimensional) allows \Ours to run fast, which is almost impossible when dealing with high-dimensional image data (e.g., 750,000-dimensional).

Third, \Ours uses the unlabeled data to train the forecasting model. As in under \textit{Design challenges}, the forecasting model forecasts how often each content category appears within a defined future time interval. This forecast is based on how frequently the content categories have appeared in the recent past.

After the offline phase, each knob configuration is characterized by the quality it achieves on different content categories as well as the profiled runtimes and cloud costs when executing the knob configuration using different task placements. When optimizing video ingestion, \Ours only considers the runtime of knob configurations together with the quality the knob configuration achieves on the current content category. This is sufficient to maximize the quality under throughput constraints and lets \Ours agnostic to the UDFs. \Ours periodically performs predictive \textit{knob planning} (e.g. every 2 days) and reactive \textit{knob switching} (e.g. every 2 seconds): Knob planning involves forecasting how often each content category appears in the future (e.g. within the next 2 days) and assigning knob configurations to the content categories based on the forecast. Knob switching involves determining the content category of the current video content and looking up what knob configuration the planning phase assigned to that category. Based on the assigned knob configuration, the available buffer space, and the profiled runtimes, \Ours then picks a knob configuration and task placement and uses it to process the next segment of video. 

\begin{figure}
    \centering
    \includegraphics[width=0.45\textwidth]{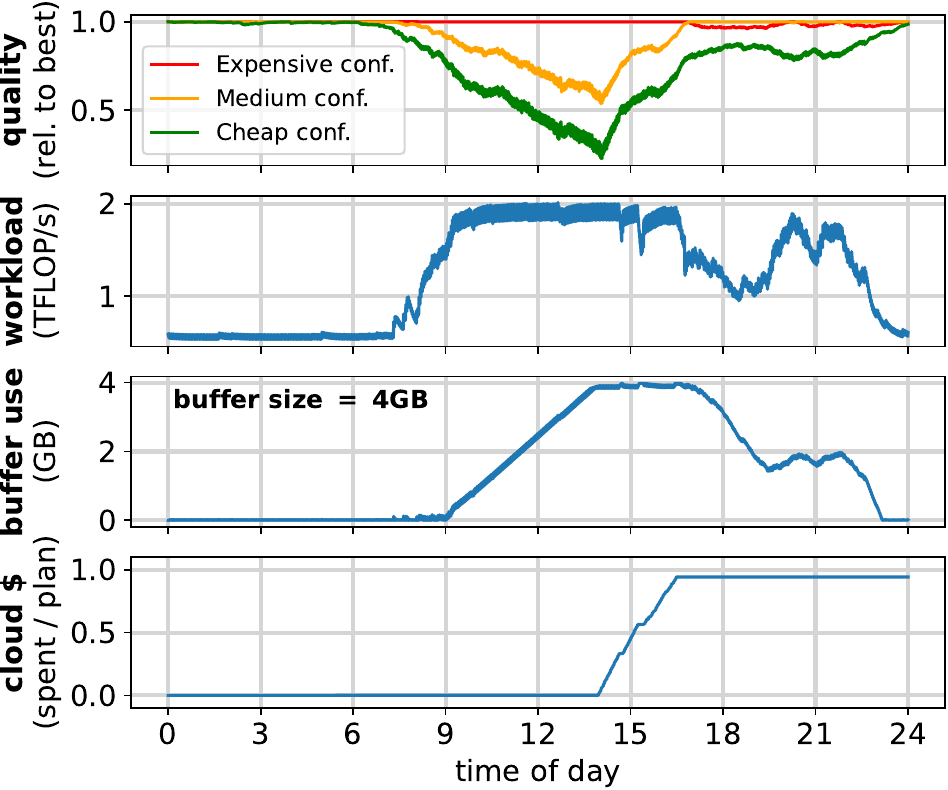}
    \vspace{-1em}
    \caption{Running the EV workload over a traffic camera.}
    \label{fig:sample-exec}
    \vspace{-1em}
\end{figure}

\textbf{\textit{Processing example}}
Figure~\ref{fig:sample-exec} shows how the knob planner and knob switcher use the provisioned resources to achieve high-quality results when running the EV example workload on 24 hours of a traffic camera stream. The uppermost plot in Figure~\ref{fig:sample-exec} shows how three different knob configurations (expensive, medium, cheap) achieve different result qualities. For the EV workload, the result quality is mainly affected by object occlusions (i.e., one car overlaps with another car). We observe that expensive configuration reliably produces high-quality results while the cheap one only produces high-quality results at night, when there is little traffic and few occlusions.

The second plot in Figure~\ref{fig:sample-exec} shows how the dynamic knob switching in \Ours causes the change in the workload (TFLOP per second). We can see that the workload is low during the night when \Ours frequently uses the cheap configurations, but high during the day when \Ours uses the expensive configurations. The data in Figure~\ref{fig:sample-exec} is smoothed and hides that \Ours switched 4500 times between knob configurations over the course of the plotted time period. If we would instead always use the most expensive configuration, the workload would be constant at 5.2 TFLOP/s.

The third plot in Figure~\ref{fig:sample-exec} shows how \Ours sets video aside into the buffer during the day when frequently running the expensive knob configuration. We can also see how \Ours catches up on processing the buffered video at 5PM, when the workload decreases. The buffer has a size of 4GB and is full at around 2 PM. When it is full, \Ours decides to offload some work to the cloud which is reflected by the rising amount of cloud credits spent in the bottom figure (note that the Y axis shows the percentage of the daily cloud budget that has been spent). We can see that \Ours's spending comes close to what it planned for that day.

\section{Offline preparation phase}
\label{sec: offline}

In the offline preparation phase, \Ours is fitted on the historical video data recorded from the same source that will be ingested in the online phase. \Ours needs a small set of labeled data (i.e., 20 minutes) and a larger set of unlabeled data (e.g., 2 weeks).
Based on this data, \Ours first leverages prior work \cite{addanki2019placeto, videostorm} to create a filtered set of knob configurations and a set of good task placements for them.
Then, \Ours clusters video content into categories allowing \Ours to reason about video content in the online phase. Furthermore, \Ours trains a forecasting model to predict the frequency that each content category appears in the near future. We describe these procedures in more detail as follows.

\subsection{Filter knob configurations and task placements}
\label{sec:knob-filter}
In order to optimize video processing while inducing little decision overheads during online ingestion, \Ours needs to decide the desirable knob configuration $k$ to process the streamed content and the placement $TP_k$ of its task graph $G_k$. Recall that the placement of $G_k$ specifies which computation components when using knob configuration $k$ to run on the cloud and which ones to run on-premises. The number of all knob configurations is exponential in the number of user-registered knobs.
Similarly, the number of all possible placements for a task graph is exponential in the number of tasks.
\Ours leverages prior work~\cite{videostorm, addanki2019placeto} to filter the set of knob configurations and task placements down to a smaller set. Thereafter, \Ours only needs to consider promising candidates in the online phase, reducing the size of the decision problem and therefore online overheads. 

We leverage the greedy hill climbing algorithm~\cite{russell2003norvig} proposed in VideoStorm~\cite{videostorm} to filter the knob configurations. We use PlaceTo~\cite{addanki2019placeto} to filter the task placements (details in Appendixs~\ref{app:filter}).

\subsection{Categorize video dynamics}
\label{sec:offline-cluster}

\Ours discretizes video content into \textit{content categories} with the property that knob configurations  achieve similar result quality for all video segments belonging to the same content category.
In this section, we describe how to identify these content categories and will discuss how to forecast them in Section~\ref{sec:offline-forecast} and
how the categories allow for efficient video ingestion in Section~\ref{sec: online}.

\Ours categorizes video content using unlabeled training data. \Ours first samples a set of video segments $\mathcal{S}'$ from the unlabeled data. \Ours then processes each segment $s\in\mathcal{S}'$ with all configurations $k\in\mathcal{K}$ and records the result quality that each $k$ achieves on the segment $s$ as $qual_s(k)$. The result quality measurement is defined by the user and will be further discussed in Section~\ref{sec: online}. We group the qualities of all configurations $k$ on a segment $s$ into a $|\mathcal{K}|$-dimensional \textit{quality vector} $qual_s = [qual_s(k_1), ..., qual_s(k_{|\mathcal{K}|})]^T$. We gather the $qual_s$ for all segments $s\in\mathcal{S}'$ to form a set of quality vectors $\mathcal{Q} = \{qual_s\,|\, s\in\mathcal{S}'\}$. Then, \Ours decides the content categories $\mathcal{C}$ by running KMeans~\cite{kmeans} on $\mathcal{Q}$.
Thereafter, the content is clustered according to the quality that the knob configurations achieve on it, ensuring that all knob configurations achieve similar result quality for the content of the same category by the property of KMeans. A content category $c\in\mathcal{C}$ is therefore characterized by a $|\mathcal{K}|$-dimensional cluster center, which denotes the average quality that the knob configurations will achieve on content belonging to $c$. We denote the cluster center as $[\accbar(k_1,c ), ... \accbar(k_{|\mathcal{K}|}, c)]$, where $\accbar(k,c)$ is the average quality that $k$ will achieve on videos categorized as $c$.

We evaluate choices for the number of categories ($k$ of KMeans) in Appendix~\ref{app:sens-kmeans} and find that \Ours is not very sensitive to $k$ as long as it is not too small (e.g. $\geq 3$). Furthermore, it is easy to tune such hyperparameters during the offline phase.

\subsection{Train the forecasting model}
\label{sec:offline-forecast}

\Ours trains a forecasting model $\mathcal{F}$ to predict how frequently each content category $c\in\mathcal{C}$ appears in the near future time interval given their frequency in the most recent history. $\mathcal{F}$ allows \Ours to effectively ration computational resources and optimally allocate them for different video content categories to come. We denote the forecasted time interval as the \textit{planned interval}.

\Ours uses a simple feed-forward neural network as forecasting model $\mathcal{F}$. We find this to be sufficient and describe its architecture in Appendix~\ref{app:eval-detail}. 
Let $r^{(T)}$ be $|\mathcal{C}|$-dimensional histogram representing the frequency each category $c\in\mathcal{C}$ appears over time interval $T$. 
The output of $\mathcal{F}$ is thus $r^{(PI)}$ where $PI$ is the planned interval. 
The input to $\mathcal{F}$ is the content histograms of the most recently ingested data. 
We split the most recent time interval $T_{input}$  into $n$ equally-sized intervals $T_{input} = [T_1, ..., T_n]$ and provide their category occurring frequency $[r^{(T_1)}, ..., r^{(T_n)}]$ as time-series inputs to $\mathcal{F}$. We evaluate choices of $T_{input}$ and $n$ in Appendix~\ref{app:sens-forecast} and find that \Ours is not very sensitive to them as long as both $T_{input}$ 
and $n$ are reasonably large (i.e. $T_{input}$ is a couple of days and is split into intervals of a couple of hours). 

\Ours pre-trains $\mathcal{F}$ in the offline phase using the unlabeled data, which we describe the detail in Appendix~\ref{app:train-data}. Furthermore, $\mathcal{F}$ can be fine-tuned in the online phase using the recently ingested data to provide more accurate forecasting.

\section{Online video ingestion}
\label{sec: online}

After completing the offline learning phase, \Ours is ready to ingest live video streams.
During live ingestion, \Ours uses both a predictive component (\emph{knob planner}) and a reactive component (\emph{knob switcher}) to make knob tuning decisions. 
The predictive knob planner periodically forecasts trends in the video content and lets \Ours make knob tuning decisions with the future workload in mind. 
This allows \Ours to put the provisioned compute resources to optimal use and prevents premature use of buffer space and cloud credits, making use of expensive knob configurations when they have the greatest impact. However, while it is possible to forecast long-term trends in the content,  the exact short-term occurrence of content is subject to substantial noise. Thus, \Ours  also uses a reactive knob switcher that switches between knob configurations based on the current content. The knob switcher presents a way to leverage the forecasted workload trends while being robust to short-term noise. 
In the following section, we describe the algorithms used for both the knob planner and the knob switcher.

\subsection{Knob planner}
\label{sec:knob-plan}

The knob planner computes a \emph{knob plan} that specifies which knob configurations $k\in\mathcal{K}$ to use for each content categories $c\in\mathcal{C}$ to maximize the overall result quality given the available compute resources. Such assignment of knob configurations to $c$ is based on the forecasted \textit{content distribution}, which specifies how frequently each knob configuration will appear over the forecasted interval. Recall from Section~\ref{sec:offline-forecast}, we refer to this interval as the the \emph{planned interval}.
We find that accurate forecasts can be achieved a couple of days into the future and consequently re-compute the knob plan every couple of days using a fresh forecast. 

Formally, the knob plan generates a histogram $\alpha_c$  over knob configurations $\mathcal{K}$ for each content category $c \in \mathcal{C}$. $\alpha_c$ determines how often a knob configuration $k \in \mathcal{K}$ should be used for processing content of category $c$ - i.e., there is one bucket in the histogram for each knob configuration, indicating the relative frequency with which that configuration should be chosen for the content category.  Let $\alpha_{k,c}$ denote the frequency that histogram $\alpha_{c}$ assigns to knob $k \in \mathcal{K}$ (i.e., how often knob $k$ should be used to process the content of category $c$). A knob plan $\mathcal{P}$ is thus defined as the set containing the histograms for all content categories: $\mathcal{P} = \{ \alpha_{c} \; | \; c \in \mathcal{C} \}$.


Finding a knob plan that maximizes the result quality under the compute budget involves jointly optimizing the histograms for all content categories. Each category's histogram determines the total resource consumption for processing content of the category, which in turn determines how many resources are available for the remaining categories. \Ours creates a knob plan in two steps.

First, the knob planner uses the pre-trained model $\mathcal{F}$ from the offline phase to forecast how often (the ratio $r_c$ described in Section~\ref{sec: offline}) each content category will appear over the planned interval. 

Second, using the forecasted content ratios $r_c$, \Ours formulates the assignment of knobs to content categories as a linear program. This allows \Ours to find the globally optimal knob plan $\mathcal{P}$. \Ours  maximizes the expected overall result quality using the content category cluster centers computed in the offline phase. As described in Section~\ref{sec: offline}, each content category $c \in \mathcal{C}$ is defined by a KMeans cluster center, which is a vector whose $i$-th element denotes the average quality $\accbar(k_i,c)$ that knob configuration $k_i$ achieves on the content of category $c$. Given the average quality of each knob configuration for each content category, the solution of the linear program maximizes the overall expected quality while being constrained by the compute budget $budget$.\footnote{The unit of the compute budget is given in $core*s$ using the on-premise server cores. \Ours internally takes care of converting the user-defined cloud credits budget.}

\vspace{-1em}
\begin{alignat}{5}
  & \text{maximize }   & \quad & \sum_{k,c}\alpha_{ k,c} * r_c * \accbar(k, c)           & \quad               &           & \label{linpro1} \\
  & \text{subject to } & \quad & \sum_{k,c} \alpha_{k,c} * r_c * cost(k) \leq budget & \quad               &           & \label{linpro3} \\
  &                    & \quad & \sum_{ k} \alpha_{k,c} = 1,    \ \ \ \alpha_{k,c} \geq 0                      & \quad \hspace{0pt} & \forall c & \label{linpro2}
\end{alignat}
\vspace{-1em}

The decision variables of the linear program are $\alpha_{k,c}$, which determine how often the content of category $c$ should be processed by configuration $k$ and thereby make up the knob plan. The goal of the knob plan is to maximize the overall result quality, which is denoted by Line~\ref{linpro1}. Line~\ref{linpro3} denotes that the total amount of cost should stay below the user-specified budget. Finally, Line~\ref{linpro2} enforces that the assigned ratios $\alpha_{k,c}$ add up to 1 for each content category (this is merely for normalization).

We use an off-the-shelf solver~\cite{scipy} which is able to find the solution to this linear program in less than a second for the problem sizes encountered by \Ours. After finding the optimal value for the decision variables $\alpha_{k,c}$, we have  the knob plan $\mathcal{P}$ which tells us how often to use each knob $k$ to process the content of category $c$ in order to achieve maximum quality given the constrained computing resources. In Section~\ref{sec:knob-switch}, we show how $\mathcal{P}$ can be leveraged to efficiently switch between knob configurations.

\subsection{Knob switcher}
\label{sec:knob-switch}

Based on the current video content, the knob switcher reactively determines which knob configuration $k_{next}\in\mathcal{K}$ to use and which tasks of $k_{next}$'s task graph $G_{k_{next}}$ to execute on the cloud and which tasks to execute on-premises. The knob switcher is designed to be lightweight and doesn't induce significant decision overheads, even when run frequently. It decides on the next knob configuration $k_{next}$ and task placement $p_{next}$ in three simple steps: First, it determines the category $c_{cur}\in\mathcal{C}$ that the current content belongs to. Second, it looks content category $c_{cur}$ up in the knob plan to obtain the configuration histogram $\alpha_{c_{cur}}$ that the knob plan assigns to $c_{cur}$. Third, the knob switcher picks knob configuration $k_{next}$ based on $\alpha_{c_{cur}}$ along with a task placement $p_{next}$ --- the knob switcher hereby guarantees to never overflow the buffer. In the following, we describe how the knob switcher performs each of these steps in more detail.

In the first step, the knob switcher determines the category $c_{cur}$ of the current content merely using the reported quality $qual^*(k_{cur})$ of the current knob configuration $k_{cur}$. This allows the knob switcher to select a category in a low overhead way, rather than running an expensive processing step on the video directly. 
Specifically, given  $qual^*(k_{cur})$, the knob switcher selects the current content category $c_{cur}$ as the one whose average quality for $k_{cur}$ ($\accbar(k_{cur}, c_{cur})$) matches the currently reported quality ($qual(k^*)$) the closest. The average quality $\accbar(k_{cur}, c)$ of $k_{cur}$ for a category $c\in\mathcal{C}$ is given by $c$'s cluster center (see Section~\ref{sec:offline-cluster}). This is denoted by Equation~\ref{equ:cat}. 

\vspace{-1em}
\begin{align}
\label{equ:cat}
c_{cur} = \underset{c\in\mathcal{C}}{\textrm{argmin}} \; \big| \accbar(k_{cur}, c) - qual^*(k_{cur}) \big|
\end{align}

Note that the knob switcher's content classification is analogous to traditional classification with KMeans but only uses one vector dimension since the other dimensions are unattainable. This works well in \Ours's case because the content of different categories will induce different result qualities for all knob configurations. As a result, the quality of one knob configuration is sufficient to discriminate between content categories. We experimentally verify this in Section~\ref{sec:microbenchmarks}. 

In the second step, the knob switcher then looks up  the derived content category $c_{cur}$ in the knob plan $\mathcal{P}$. This yields a  histogram $\alpha_{c_{cur}}$ dictating how often each knob configuration $k\in\mathcal{K}$ should be used to process the content of the current category $c_{cur}$:

In the third step, the knob switcher  determines the  knob configuration $k_{next}$ that will be used for processing the newly arriving content, together with task placement $p_{next}$ that determines which tasks of $k_{next}$'s task graph to execute on the cloud and which ones to execute on-premises. The knob switcher tries to adhere as closely to the planned histogram $\alpha_{c_{cur}}$ as possible and therefore keeps a histogram $\widehat{\alpha}_{c}$ for each $c\in\mathcal{C}$, which denotes how frequently each knob configuration has actually been used to process the content of category $c$. To adhere as closely to the knob plan as possible, the knob switcher picks the knob configuration $k_{next}$ that minimizes the difference between  $\widehat{\alpha}_{c_{cur}}$ and $\alpha_{c_{cur}}$. This is denoted by Equation~\ref{equ:pickk}. 
Finally, the knob switcher picks a placement $p_{next}$ for $k_{next}$. \Ours picks the cheapest placement of $G_{k_{next}}$ that does not overflow the buffer. 

\vspace{-1em}
\begin{align}
\label{equ:pickk}
k_{next} = k_i \;\textrm{ with } i = \underset{1\leq i\leq|\mathcal{K}|}{\textrm{argmax}} \, (\alpha_{c_{cur}}[i] - \widehat{\alpha}_{c_{cur}}[i])
\end{align}
\vspace{-1em}

It is worth noting that there is an edge case when picking the task placement $p_{next}$: Some knob configurations do not possess task placements that run in real-time, even when heavily adding cloud compute. Reasons for this include limited bandwidth to the cloud, high round trip times to the cloud, and limited opportunities for adding parallelism to the DAG execution. 
If all placements of $k_{next}$ would make \Ours's buffer overflow, the knob switcher will choose a different configuration $k_{next}'$ to be the next one. This knob configuration $k_{next}'$ is the next less qualitative one compared to $k_{next}$. Like for $k_{next}$, the knob switcher will pick the cheapest placement of $k_{next}'$ that does not overflow the buffer. If all placements of $k_{next}'$ would overflow the buffer, the knob switcher will recursively apply this procedure of picking the next less qualitative knob configuration until it finds a configuration and task placement that do not overflow the buffer.

In summary, the knob switcher uses three steps to find a knob configuration $k_{next}\in\mathcal{K}$ along with a task placement $p_{next}$ while adding little runtime overheads to the ingestion process. The knob switcher tries to adhere as closely to the knob plan $\mathcal{P}$ as possible, only deviating from the knob plan when this is required to avoid a buffer overflow. This ensures that the knob switcher maximizes the result quality with the given resources.

\section{Evaluation}
\label{sec:eval}

We evaluate \Ours on several real-world applications, covering public health monitoring, traffic planning, and social media analysis. We describe these workloads in subsection~\ref{sec:eval-workloads}. Then, we evaluate \Ours on the following aspects:

\begin{itemize}
    \item[\S\ref{sec:eval-cost}] What cost savings does \Ours achieve versus using a static knob configuration?
    \item[\S\ref{sec:eval-ablation}] How much do cloud bursting and buffering individually contribute to cost savings in different quality regimes? When do they perform well and when don't they?
    \item[\S\ref{sec:eval-overheads}] How much decision overhead does \Ours impose at different scales?
    \item[\S\ref{sec:microbenchmarks}] How accurate are knob planner and knob switcher, and what effect do inaccuracies have on \Ours's end-to-end performance?
\end{itemize}

We further evaluate different hyperparameter choices of \Ours in Appendix~\ref{app:hyperparams} (e.g., number of content categories (KMeans clusters), periodicity of running the knob switcher, and more). We hereby find that \Ours's end-to-end performance is insensitive to many of the hyperparameters as long as they are within a reasonable range.





\subsection{Implementation}
\label{sec:eval-impl}

We implement \Ours in Python on top of Ray~\cite{ray}. We instantiate several Ray actors for both the on-premise and the cloud version of each UDF. The number of duplicate actors is based on the number of logical cores of the machine. We only map UDFs to Ray actors; all of \Ours's components run in the parent process and synchronize the calls to the actors. We discuss implementation choices in more detail in Appendix~\ref{app_impl_details}.

We use AWS Lambda~\cite{aws-lambda} to run UDFs in the cloud and provision 3GB of memory for each cloud function. 
To simulate incoming video streams in real time, we read video frames from the disk and pause appropriately between frames to guarantee 30 fps streaming rate. All workloads are compute-bound and we find in Appendix, that in our experiments decode only amounted to 5\% of the overall runtime. The streamed video is encoded in H.264~\cite{h264} and has a resolution of 1280 × 720 (HD). In our experiments, each frame is decoded when arriving in the system (as part of the user code).

When sending full or partial frames to the cloud, we compress them to JPEG-1 format~\cite{jpeg}. We then serialize the JPEG using Base64~\cite{base64} and send the string as part of an HTTPS request. The overhead for encoding and decoding is negligible compared to the transfer time saved through compression.

\begin{figure*}
    \centering
    \vspace{-1.2em}
    \includegraphics[width=0.9\textwidth]{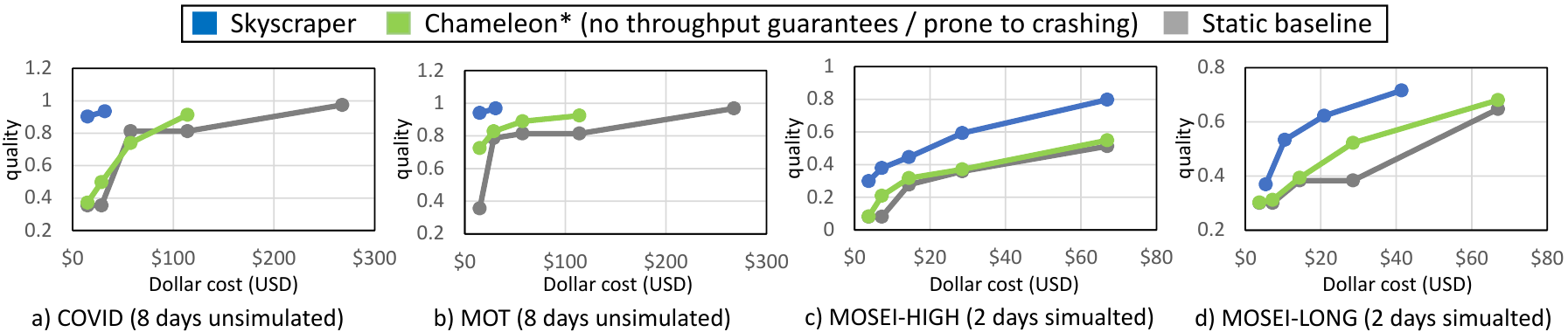}
    \vspace{-1em}
    \caption{Cost-quality trade-off of \Ours, Chameleon$^{*}$ and statically using the same knob throughout ingestion.}
    \label{fig:chameleon}
    \vspace{-1em}
\end{figure*}

\subsection{Workloads}
\label{sec:eval-workloads}

We evaluate \Ours using three workloads on public health monitoring, traffic planning, and social media analysis. They cover a diverse set of computer vision primitives including object detectors, trackers, and classifiers, as described below. 

\textbf{\textit{COVID-19 safety measures (COVID)}} During the coronavirus pandemic, decision-makers have executed several safety measures to slow down the spread of the virus. Such measures include wearing facial masks and social distancing. Measuring where and how strictly people adhere to these measures can be used for decision-making and informing people at risk. 
The COVID workload consists of a YOLOv5 object detector~\cite{yolo} to detect pedestrians and a KCF tracker~\cite{kcf} to track the detected pedestrians ("detect-to-track"). After the detection, for each detected pedestrian, the workload employs homography~\cite{homography} to measure the pedestrian's distance from others. 

This workload contains the following knobs: 1) \textit{frame rate} at which video is processed (\{30FPS, 15FPS, 10FPS, 5FPS, 1FPS\}), 2) \textit{object detection rate} to run object detector (every \{1, 5, 30, 60\} frames) and 3)  \textit{tiling for object detection} that slices the frames into (\{1x1, 2x2\}) tiles. The detailed semantics of these knobs are provided in Appendix~\ref{sec:app-workloads}.

The workload is executed on an 8-day video stream of a busy shopping street in Tokyo.\footnote{The Koen-Dori street in the Shibuya district: \url{https://youtu.be/gALQR-nsEME} }
We measure quality in terms of the number of people detected and tracked over time as YOLO has a low false positive rate and KCF trackers reliably report tracking errors.

\textbf{\textit{Multi-object tracking (MOT)}} Multi-object tracking (MOT) is a key primitive in many video analytical pipelines. In this workload, we adopt the recent state-of-the-art TransMOT~\cite{transmot} tracker on MOT benchmark~\cite{MOT20} and introduce four tunable knobs: 1) \textit{frame rate} (every \{1, 5, 30, 60\} frames), 2) \textit{number of tiles} (\{1x1, 2x2\} tiling), 3) \textit{length of history} denoting the number of historical frames (\{1, 2, 3, 5\}) as the TransMOT input, and 4) \textit{model size} (\{small, medium, large\}) that specifies different parameter sizes of the pre-trained TransMOT.
The details of TransMOT and its tunable knobs are provided in Appendix~\ref{sec:app-workloads}.

We run MOT on a stream of a traffic intersection, Shibuya in Tokyo to track pedestrians for 8 days.
MOT's processing quality is defined as the sum of tracked pedestrians weighted by the model's reported certainty. With this quality metric, we want to evaluate how \Ours maximizes model certainty as a proxy for accuracy as proposed in prior work~\cite{pred-certain1, pred-certain2}.

\textbf{\textit{Multi-modal opinion sentiment and emotion 
intensity (MOSEI)}}
This workload is synthetic and simulates a video stream analysis application on Twitch. The number of incoming streams varies over time and mimics the number of live Twitch streams over two days.\footnote{As recorded by Twitch Tracker at \url{https://twitchtracker.com/statistics/active-streamers}} 
We further introduce two types of spikes to evaluate \Ours under difficult conditions:

$\bullet$ \textit{MOSEI-HIGH}: We introduce high but short peaks in workload, consisting of 62 concurrently incoming video streams. This makes cloud bursting difficult due to bandwidth limitations.

$\bullet$ \textit{MOSEI-LONG}: We introduce a long peak of continuous workload. In this case, the buffer alone cannot handle all the extra work. 

We use the CMU-MOSEI~\cite{cmu-mosei-2018-multimodal} dataset to simulate incoming video streams, as it has ground truth labels that allow us to train the models used in the workload.
It contains various talking head videos from YouTube.
The task of the MOSEI workload is to classify the opinion sentiment of the speaker using both the audio and the visual content. 
CMU-MOSEI provides extracted features from the video with ground-truth labels. 
We trained a neural network on CMU-MOSEI's training set and used its test set to evaluate \Ours. 

MOSEI workload contains the four knobs: 1) \textit{frame rate}, 2) \textit{frequency of sentiment analysis} that we may run sentimental analysis model once every \{1, 2, 3, 4, 5, 6, 7\} sentences of the spoken audio and video, 3) \textit{model size} of the sentimental analysis model, and 4) the \textit{number of streams} to analyze.

Further details about this dataset, the entity extraction DAG, and the tunable knobs are presented in Section~\ref{sec:app-workloads} in the Appendix.
We evaluate the processing quality as the weighted sum over the ingested streams weighted on model's reported certainty.

Due to space limitation, we describe the hyperparameters of \Ours for all four workloads in Appendix~\ref{app:hyperparams}.


\subsection{Cost efficiency}
\label{sec:eval-cost}

In this section, we evaluate the end-to-end cost savings that \Ours achieves on these workloads. We hereby compare \Ours to two baselines. The Static baseline processes the video streams statically using the same knob configuration throughout the stream. The Chameleon* baseline refers to an adapted version of Chameleon~\cite{chameleon}. We equip Chameleon with a buffer and adapt it to set video aside when the provisioned hardware cannot process it in real-time. This allows Chameleon to achieve cost savings, since it doesn't need to be provisioned to handle peak workload. However, Chameleon* is not practical and may easily crash, as its lack of throughput guarantees may lead to buffer overflows. We benchmarked Chameleon* on several hardware setups and only report the numbers where it didn't crash during the benchmark.

For each system, we report the overall result quality that the system achieves on different hardware set ups.
Since we do not have access to a wide range of compute servers, we use Google Cloud VM instances as the provisioned, always-on hardware ("on-premise servers"). 
In the case of \Ours, which additionally uses AWS Lambda, we have verified that the bandwidth and latencies from the Google Cloud VMs to AWS Lambda realistically reflect the ones of commodity on-premise setups. In our experiments, we consider the following Google Cloud machines:

$\bullet$ \verb|e2-standard-4|: 4 vCPUs, 16 GB memory, 0.14 USD/h

$\bullet$ \verb|e2-standard-8|: 8 vCPUs, 32 GB memory, 0.27 USD/h

$\bullet$ \verb|e2-standard-16|: 16 vCPUs, 64 GB memory, 0.54 USD/h 

$\bullet$ \verb|e2-standard-32|: 32 vCPUs, 128 GB memory, 1.07 USD/h

$\bullet$ \verb|c2-standard-60|: 60 vCPUs, 240 GB memory, 2.51 USD/h

While these instance types do not possess hardware accelerators (e.g., GPUs), we note that there is nothing fundamental about \Ours that would prevent users from using hardware different from only CPUs. If a user provisions \Ours with a server that contains hardware accelerators, the application's UDFs would need to make sure that the hardware accelerators are used when executing the UDF. In the offline phase, \Ours will then just measure the UDF's runtime and work normally without any modifications.

Figure~\ref{fig:chameleon} visualizes the cost of processing the workloads from Section~\ref{sec:eval-workloads} with each system. On average, content category changes occured every 42s for COVID, every 43s for MOT, every 30s for MOSEI HIGH, every 24s for MOSEI LONG. However, all workloads had some periods with very frequent category changes and others with few category changes. Table~\ref{tab:chameleon} in Appendix~\ref{app:chameleon} further provides the numeric measurements depicted in Figure~\ref{fig:chameleon}. The total cost of each system is derived from the cost of renting the cloud hardware. In Appendix~\ref{app:cloudcost}, we estimate that the same amount of computing costs 1.8$\times$ more when using a Google Cloud VM than when using a provisioned on-premise server (this estimate is high and in favor of the baselines). Thus, the total cost of all systems is given by the cost of renting the Google Cloud VMs divided by $1.8$ plus the cost of the AWS Lambda workers.

\smallskip
\textit{\textbf{Summary.}} Overall, \Ours offers significantly better cost-quality trade-offs than current approaches. \Ours's performance benefits are especially large on the MOT workload: \Ours is $8.7\times$ cheaper than the static baseline at a comparable quality. Furthermore, \Ours is $3.7\times$ cheaper than Chameleon* at a better quality. Chameleon* suffered from large profiling overheads. For the COVID and MOT workload, our results are comparable to what the authors report in the Chameleon paper (2-3$\times$ speedup over the static baseline at the highest quality level). For the MOSEI workloads, the profiling overheads were especially large since the expensive knob configurations cause large amounts of work.

\blfootnote{* Chameleon* is an adapted version of Chameleon~\cite{chameleon} that uses a buffer. However, Chameleon* is not practical and would frequently crash in practice due to overflows of the unmanaged buffer.}


\vspace{-1em}
\subsection{Ablation study}
\label{sec:eval-ablation}

\begin{figure*}[t]
\vspace{-1.5em}
  \includegraphics[width=0.88\textwidth]{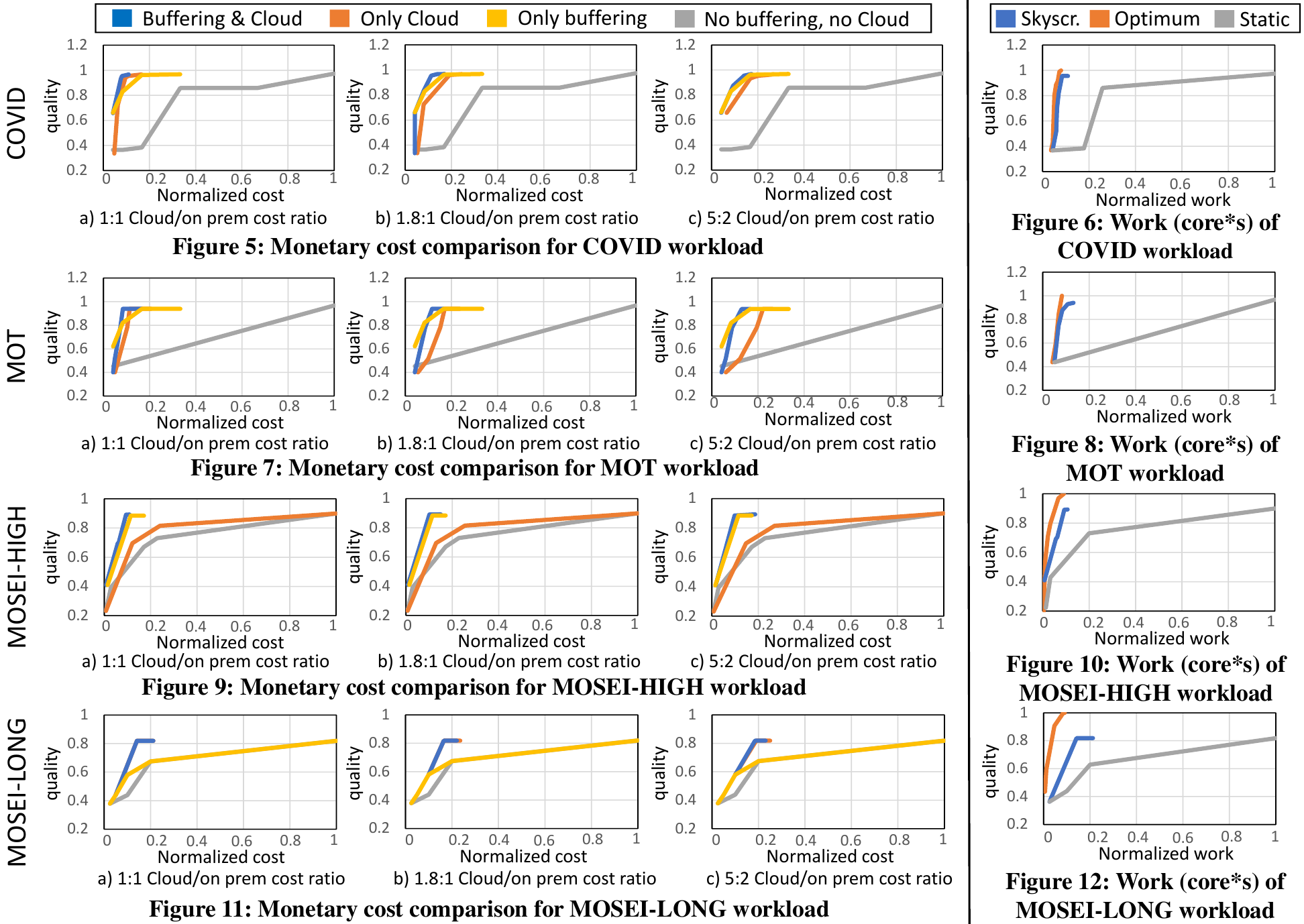}
\end{figure*}
\setcounter{figure}{12}

To evaluate how much buffering and cloud bursting individually contribute to the cost savings, we run an ablation study where we independently disable them. 
Running this ablation study on unsimulated hardware is prohibitively expensive (i.e., we need to conduct dozens of measurements as the one in Figure~\ref{fig:chameleon}), so we can only afford to analyze with simulated results. We use a simple but accurate simulator, that we describe in Appendix~\ref{app:sim-desc}. We evaluate the accuracy of the simulator on the benchmarked workloads in Appendix~\ref{app:eval-simulator} and find that it is reasonably accurate.

We use two metrics to evaluate the performance of \Ours:

(1) \textbf{The monetary cost} of processing the workload. We hereby also evaluate \Ours for different cost ratios between the on-premise and the cloud computing. In Appendix~\ref{app:cloudcost}, we estimate that a ratio of 1:1.8 between on-premises and AWS Lambda is realistic at the current market prices. 
When evaluating the monetary cost, we evaluate four variations of \Ours:

(1a) \emph{No buffering, no cloud:} We disable both buffering and cloud bursting. Effectively, this corresponds to not switching knob configurations and only using the most qualitative knob configuration that runs in real time on the given on-premise server.

(1b) \emph{Only buffering:} \Ours may only use placements that place every task on-premise and can not use the cloud. 

(1c) \emph{Only cloud:} \Ours may use the cloud but not buffering. 

(1d) \emph{Buffering \& cloud:} This corresponds to standard \Ours.

(2) \textbf{The amount of work} measured in $core*second$s used in the processing. This is independent of whether the computation is buffered or executed on the cloud or on premises.
When evaluating the amount of work, we compare \Ours to two baselines:

(2a) \emph{Static:} This baseline corresponds to statically using the same knob configuration. It is similar to baseline (1a) where \Ours also statically uses the same configuration.

(2b) \Ours: We measure the amount of work that \Ours performs for processing the workload.

(2c) \emph{Optimum:} The optimum baseline fully leverages the ground truth to always choose the optimal knob configuration. Specifically, given the performance of each knob configuration beforehand, it uses the greedy 0-1 knapsack approximation to choose knob configurations that maximize quality under certain budget.

Figures 6, 8, 10, 12 show the cost-quality trade-off curves for the COVID, MOT, MOSEI-HIGH, and MOSEI-LONG workloads. Figures 7, 9, 11, 13 show the work-quality trade-off curves.

For the COVID and MOT workload, \emph{Only cloud} and \emph{Only buffering} alone can achieve significant speed-ups over the baseline.
For both workloads, when combining the two (\emph{Buffering \& cloud}), peak quality can be roughly reached at $1.5\times$ less cost than when only buffering or only using the cloud for a cost ratio of 1.8:1.
For 5:2 cost ratio, \emph{Only cloud} performs significantly worse, because off-loading work off to the cloud incurs a very high cost.
For 1:1 cost ratio, \emph{Only cloud} matches the performance of \emph{Buffering \& cloud} as using cloud resources has the same cost the on-premises computations.

For the MOSEI workloads, we can see how \emph{Only buffering} and \emph{Only cloud} struggle to deliver good performance for MOSEI-HIGH and MOSEI-LONG, respectively. However, we observe that \emph{Buffering \& cloud} delivers good performance on both. The reason for the bad performance of \emph{Only cloud} on MOSEI-HIGH is bandwidth limitations that limit the number of social media streams that can be offloaded to the cloud. The reason for the bad performance of  \emph{Only buffering} on MOSEI-LONG is that the buffer gets filled early on, which prevents \Ours from using expensive knob configurations for the remaining duration of the long workload peak.

Finally, Figures 7, 9, 11 show that \Ours's work reduction method performs astonishingly close to optimum. Only for the MOSEI-LONG workload (Figure 13) does \Ours leave large room for improvement.

\smallskip
\textit{\textbf{Summary.}} To certain extent, the buffering and cloud bursting optimizations are complementary to each other.
Specifically, the performance improvement of using both over using one of them is not as large as performance difference between them. Therefore, cloud bursting lessens the need for buffering and vice versa. 
However, \Ours can still achieve 1.5$\times$ cost savings in the COVID and MOT workloads over only one of the two methods. Furthermore, the MOSEI workloads show that buffering and cloud bursting struggle for different kinds of workload spikes. By combining the two, \Ours can achieve good performance for both kinds of spikes.

\subsection{Runtime overheads}
\label{sec:eval-overheads}

Appendix~\ref{app:offline_scale} evaluates the runtime of the offline phase. For the COVID workload, the overall runtime was 1.6 hours on two c2-standard-60 machines. 83\% of the time was spent creating the training data for the forecasting model, which is embarrassingly parallel and can be sped up when using more machines.

\Ours's \emph{knob planner} and \emph{knob switcher} add overheads to the online execution time. In this section, we evaluate their runtimes for different amounts of placements, content categories, and knob configurations. All runtime measurements are performed on a single core of the Intel Xeon Gold 6130 CPU running at 2.10GHz.

The worst-case runtime of the \emph{knob switcher} is linear in the total number of placements (for all knob configurations). This worst case is achieved when the knob switcher needs to iterate through all configuration-placement pairs until it finds one that does not overflow the buffer (see Section~ \ref{sec: online}).
The left plot in Figure~\ref{fig:overheads} shows the worst-case runtime as the dashed line and the average runtimes of the \emph{knob switcher} for the COVID, MOT, and MOSEI experiments.

The \emph{knob planner} conducts an inference pass through a small neural network and solves a linear program. For the linear program, the number of variables is $|\mathcal{C}|*|\mathcal{K}|$ and the number of constraints is $1+2*|\mathcal{C}|$, where $\mathcal{C}$ denotes the number of content categories and $\mathcal{K}$ is the number of knob configurations. The right image in Figure~\ref{fig:overheads} uses the heat map to visualize the overheads caused by the \emph{knob planner} for different amounts of content categories and knob configurations. This image also shows the actual runtime of \emph{knob planner} on the three workloads.

\begin{figure}[t]
    \includegraphics[width=0.4\textwidth]{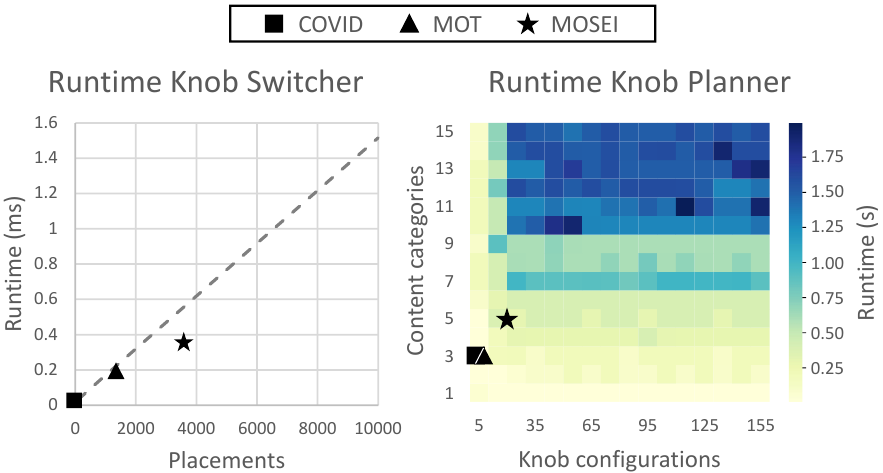}
    \vspace{-1em}
    \caption{Overheads: knob switcher ($<$1ms) and planner ($<$1s)}
    \vspace{-1.5em}
    \label{fig:overheads}
\end{figure}

\smallskip
\textit{\textbf{Summary.}} For common problem such as the COVID, MOT, and MOSEI workloads, the overheads of both the \emph{knob switcher} and \emph{knob planner} are negligible. While the \emph{knob switcher} runs every few seconds, its runtime is typically below a millisecond. Similarly, the \emph{knob planner} typically runs every few days but with a runtime below a second. We also show that the runtime overhead of our optimization is reasonable for more complicated workloads.

\subsection{Microbenchmarks}
\label{sec:microbenchmarks}

This subsection evaluates how accurately \Ours's forecasting model $\mathcal{F}$ can predict the future content distribution and how sensitive \Ours's performance is to inaccuracies in the forecast. Similarly, the subsection evaluates the accuracy at which the knob switcher classifies the video content into a content category $c\in\mathcal{C}$ and how sensitive \Ours's performance is to misclassifications. In our evaluation, we focus on the real-world workloads COVID and MOT. The MOSEI workloads are synthetically created by inducing workload spiking patterns as described in Section~\ref{sec:eval-workloads}. While these workloads present especially difficult spiking patterns for buffering and cloud bursting, the forecasting model achieves 100\% accuracy and the knob switcher particularly high performance due to the regularity and smoothness of their workload peaks. We therefore do not evaluate them in terms of accuracy in this subsection.

\textit{\textbf{Forecasting model}} We evaluate the forecasting model on 8 days of test data after training it on 16 days of unlabeled training data. We train and evaluate the forecasting model on four different lengths of the planned interval: \{1, 2, 4, 8\} days. As described in Section~\ref{sec:knob-plan}, the length of the planned interval determines the frequency of running knob planner and how long $\mathcal{F}$ needs to forecast into the future. 

We find that for both workloads, \Ours's forecasting method achieves a low Mean Absolute Error (MAE) when forecasting 1 to 4 days into the future. We denote the MAE values in Appendix~\ref{app:sens-forecast}. For both workloads, the lowest MAE was achieved when forecasting 2 days into the future, while the largest MAE was incurred when doing so for 8 days.

There is a sweet spot on how far to forecast into the future but this sweet spot is unrelated to the frequency of content category changes. Forecasting over very large time intervals is hard because events far in the future become increasingly uncorrelated to the current events, which the forecast is based on. On the other hand, forecasting over too short time periods is also hard: The streamed video content is always subject to a certain amount of randomness (e.g. a large group of people randomly walking past a camera). Over large enough time intervals, this randomness is smoothed out, which makes the forecast more precise. When this smoothing effect is not achieved, errors due to unforeseen randomness will be noticeable in the MAE of the predictions.
The high MAE when forecasting 8 days into the future shows that forecasting far into the future is difficult as events become increasingly uncorrelated to the current events, which the forecast is based on. On the other hand, forecasting over too short time periods also leads to higher MAEs: Streamed video content is always subject to a certain amount of randomness (e.g. a large group of people randomly walking past a camera). Over large enough time intervals, this randomness is smoothed out and therefore doesn't show in the MAE, which doesn't occur for forecasts over short periods.

\begin{figure}[]
    \centering
    \includegraphics[width=0.42\textwidth]{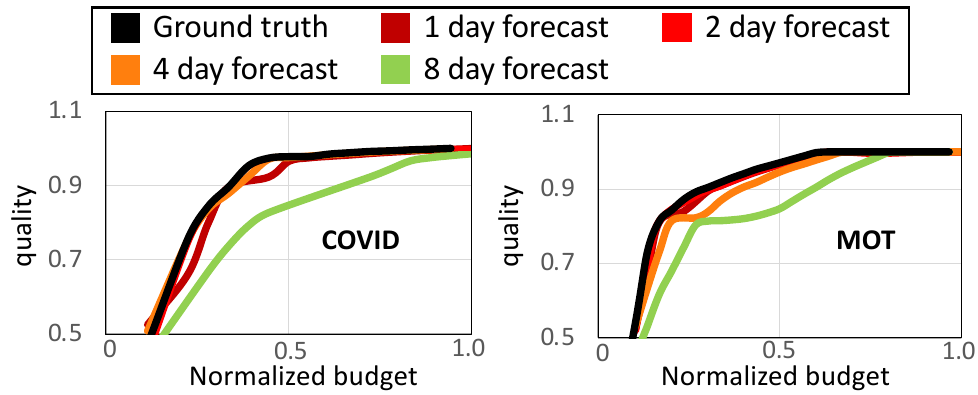}
    \vspace{-1em}
    \caption{The effect of different planned interval lengths on \Ours's end-to-end performance}
    \vspace{-1em}
    \label{fig:eval-forecast}
\end{figure}

Figure~\ref{fig:eval-forecast} shows the impact of the prediction errors in terms of end-to-end performance. For comparison, we additionally run \Ours using the ground truth content distributions instead of forecasting.
For planned interval lengths between 1 and 4, \Ours's performance is very close to the optimal performance using the ground truth predictions. However, for both workloads \Ours performs significantly worse for a planned interval length of 8.

\textit{\textbf{Knob switcher}} As described in Section~\ref{sec:knob-switch}, it is possible that the knob switcher misclassifies video content into the wrong content category. We identify two reasons for such misclassifications. First, the knob switcher classifies content based on the quality of one knob configuration. This corresponds to KMeans classification, where a vector is classified using only one dimension instead of all.
We denote misclassifications, that occur because of this as \emph{Type-A errors}.
Second, the knob switcher determines the current content category based on the past couple of seconds of the video. It will then switch to a knob configuration that is used for processing the next couple of seconds of video, which creates a time mismatch. The last couple of seconds might belong to a different content category than the next couple of seconds. We denote errors caused by this timing mismatch as \emph{Type-B errors}. Distinguishing between these two errors lets us gain insight into where performance losses come from, which could be used for further enhancements of \Ours.

In Figure~\ref{fig:eval-switcher}, we denote the standard knob switcher as described in Section~\ref{sec:knob-switch} as \emph{Standard} and compare it against two baselines: \emph{Ground truth} denoting \Ours using the ground truth content categories and \emph{No Type-B errors} denotes a baseline that partially uses the ground truth to eliminate errors of Type-B. Specifically, it determines the content category using \Ours's standard approach but on the data of a future couple of seconds (i.e., it knows how the current knob configuration would perform in the next couple of seconds without executing it). Like this, only errors of Type-A impede the performance of the \emph{No Type-B errors} baseline, which shows their impact on \Ours's end-to-end performance.


Figure~\ref{fig:eval-switcher} shows that the knob switcher's misclassifications have a negative impact on \Ours's end-to-end performance when using the \emph{Standard}. The misclassification rate of \emph{Standard} is 2.1\% on COVID and 6.6\% on the MOT workload.
However, the performance of the \emph{No Type-B errors} baseline almost matches the \emph{optimum}. This suggests that the remaining Type-A errors barely impede the overall performance. These errors constitute  0.5\% of the knob switcher's error rate on COVID and 3.7\% on the MOT workload.

\begin{figure}[]
    \centering
    \includegraphics[width=0.42\textwidth]{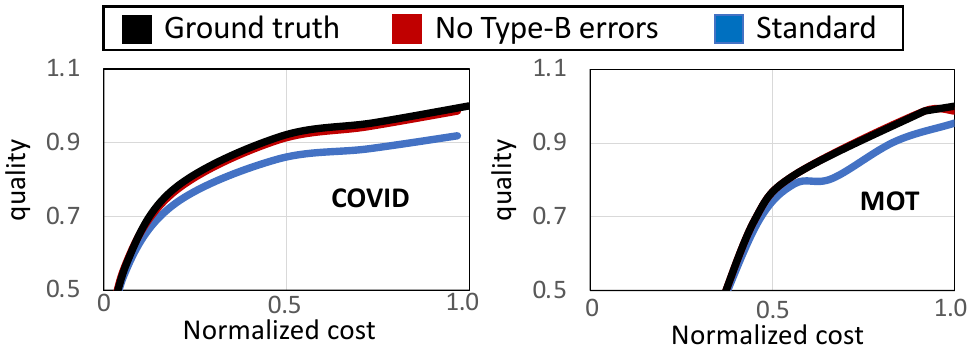}
    \vspace{-1em}
    \caption{End-to-end performance of knob switcher against baselines that leverage ground truth for content classification}
    \vspace{-1.7em}
    \label{fig:eval-switcher}
\end{figure}

\textit{\textbf{Summary}} The microbenchmarks provide two insights. First, when forecasting between 1 to 4 days into the future, \Ours's forecasting method is accurate and does not significantly harm end-to-end performance when compared to using the ground truth as forecast. However, when forecasting further into the future (e.g., 8 days), the forecasts become less accurate, which shows an effect on \Ours's end-to-end performance. Second, misclassifications of the knob switcher negatively impact \Ours's performance. We hereby identify a time mismatch as the sole driver for the performance losses. This timing mismatch occurs because the knob configuration to process the next couple of seconds with is based on the content of the last couple of seconds.
\section{Related Work} \label{sec:rel}


The cost problem of video processing has previously been recognized~\cite{romero2021llama, romero3}. 
While we are not aware of past research which manages video streams like in a data warehouse, several systems propose end-to-end solutions for managing archived collections of video like in a relational database system~\cite{seiden, eva-demo, eva-sigmod,haynes2021vss, viva, daum2021tasm, xu2019vstore, xu2022eva, lu2016optasia}. Likewise, we are not aware of past work that directly addresses the \vetl problem, but there are several lines of work on efficient video processing that are relevant to \Ours.

\textbf{\textit{Content-adaptive knob tuning systems.}}
Content-adaptive knob tuning systems aim at saving computational work by dynamically adjusting knobs that are inherent to CV workloads to the video stream's content.
Chameleon performs content-adaptive knob tuning for general CV workloads~\cite{chameleon}.
However, Chameleon assumes that each knob configuration can be run in real-time on the provisioned hardware resources ("peak provisioning"). Chameleon then minimizes the average processing time per frame. As discussed in Section~\ref{sec:introduction}, such systems cannot deliver cost savings while also adhering to throughput guarantees, which is required in the \vetl problem.
Zeus is another content-adaptive knob tuning  system~\cite{zeus}, but cannot be used for general-purpose \vetl, as it is specific to action detection (e.g., detect someone crossing the street).

\textbf{\textit{Query-load-adaptive knob tuning systems.}}
Instead of adapting to the streamed content, some systems tune the knobs of a CV workload solely based on the concurrently running queries (while being agnostic to the streamed content). These systems are useful in scenarios where users issue dynamic queries over video streams, which require the system to dynamically multiplex compute resources among the queries. VideoStorm~\cite{videostorm} and VideoEdge~\cite{videoedge} go beyond dynamic resource allocation and also tune the queries' knobs based on the other queries that are concurrently running. However, in scenarios where the query load remains static, there is no benefit in dynamically adapting to the query load. In \vetl,
a constant set of jobs is used to ingest the video streams. In contrast to VideoStorm and VideoEdge, \Ours therefore dynamically adapts to changes in the video content instead of the query load.

\textbf{\textit{Streaming ETL.}} Treating data warehouse ingestion as a stateful stream processing problem is an established approach~\cite{golab}, which is successfully used in many big data applications \cite{meehan}. 
Like \Ours, traditional streaming ETL is also concerned with maintaining data quality while handling fluctuating workloads 
without peak provisioning.
This is typically achieved through methods like back pressure or load
shedding, which mitigate workload peaks arising from fluctuating volumes of arriving data~\cite{tatbul}. However, in \vetl, data often arrives at constant volume, and only the content of the data changes. In contrast to traditional streaming systems, \Ours’s optimizations therefore focus on adapting to the content of the streamed data and not to its volume.

\textbf{\textit{General-purpose cloud offloading.}} Several works have previously explored the idea of offloading work from an on-premise server to on-demand cloud workers~\cite{cloud1, cloud2, cloud3, cloud4, cloud5, cloud6, cloud7, cloud8}. These works assume that jobs occasionally arrive and these jobs may be executed locally or offloaded to the cloud. However, these works only optimize the placement of work and do not reduce work by means like knob tuning, which is \Ours's main optimization. 

\textbf{\textit{Task-specific computer vision optimizations.}}
Several works optimize the application of CV for specific tasks and queries. While these methods cannot be used to optimize arbitrary \vetl jobs, they can be used inside \Ours's UDFs to further reduce cost. General methods to improve the efficiency of neural networks include model compression \cite{compr2, compr1}, compact neural architectures, \cite{compact1,compact2,yolo}, and knowledge distillation \cite{distill3, distill1, distill2, noscope}. 
Further works propose efficient CV primitives that are query-aware or content-adaptive~\cite{miris, noscope, thia, otif, evol-boxes, tahoma, live-spatula}. Finally, some works reduce processing costs of certain video queries by intelligently skipping frames~\cite{probabilistic-predicates, core, abae, blazelt, exsample, eko, tasti, voodoo,focus-2018}.

\section{Conclusion} \label{sec:conc}

In this paper, we defined the problem of \vetl for transforming video streams to a queryable format through expensive ML-based video processing DAGs. In response, we introduced \Ours, which uses content-adaptive knob tuning to reduce the cost of the \vetl Transform step while adhering to \vetl's throughput requirements on constrained hardware resources. \Ours supports conversions to arbitrary query formats. 




\begin{acks}
 We thank the Data Systems and Artificial Intelligence Lab (DSAIL) for supporting this work.
\end{acks}

\clearpage

\bibliographystyle{ACM-Reference-Format}
\bibliography{sample}

\clearpage

\appendix

\section{Filtering the knob configurations and task placements}
\label{app:filter}

\subsection{Filter knob configurations}

Processing video with a knob configuration $k$ incurs an amount of work (i.e. FLOPs) and a processing quality determined by $k$. However, not all knob configurations are pareto-optimal: one configuration may achieve worse result quality while incurring more work than another. \Ours proceeds in two steps to create a filtered set $\mathcal{K}$ of knob configurations that lie on an (approximated) work-quality pareto frontier.

\Ours samples a set $\mathcal{S}$ of video segments from the unlabeled training data and then runs greedy hill climbing on the sampled segments as proposed in VideoStorm~\cite{videostorm}. In the following, we describe \Ours's sampling method that yields $\mathcal{S}$. The goal of the sampling method is to sample $n_{search}$ segments that have widely different content dynamics (e.g. one with many cars, one with few cars). Given a small set of labeled data, and a larger set of unlabeled data, as well as the hyperparameter $n_{search}$, which is set by the user,  \Ours proceeds in two steps to compute $\mathcal{S}$.

First, \Ours finds the cheapest knob configuration $k^-$ and the most qualitative knob configuration $k^+$. Both of these configurations are guaranteed to lie on the work-quality pareto frontier. The cheapest configuration $k^-$ can simply be found by measuring the runtimes of all knob configurations on the on premise cluster. The most qualitative configuration $k^+$ can be found by running all configurations on the labeled training data and picking the one that achieves the best accuracy.

Second, \Ours samples $n_{pre}$ segments $\mathcal{S}_{pre}$ from the unlabeled training data uniformly at random and processes all of them with $k^-$ and $k^+$. The quality that $k^-$ and $k^+$ achieve on a segment are recorded as 2-dimensional \textit{quality vectors}. Let $\mathcal{Q}_{pre}$ denote the set of quality vectors for the segments in $\mathcal{S}_{pre}$. \Ours now greedily selects segments from $\mathcal{S}_{pre}$ and adds them to the initially empty set $\mathcal{S}$. \Ours first picks the segment $s\in\mathcal{S}_{pre}$ with the smallest L2 norm and adds it to $\mathcal{S}$. Then, \Ours iteratively adds the segment $s'\in\mathcal{S}_{pre}$ to $\mathcal{S}$ such that the newly added segment $s'$ is the one which differs the most from all the previously added segments in $\mathcal{S}$. The segment $s'\in\mathcal{S}_{pre}$ to be added is the segment with the largest distance to its closest element in 
$\mathcal{S}$ 
(i.e. $s' = \underset{s\in\mathcal{S}_{pre}}{\textrm{argmax}} (\min(\{||s'-s''||\;|\,s''\in\mathcal{S}\}))$ ). 

After $n_{search}-1$ iterations, $\mathcal{S}$ is a set of $n_{search}$ video segments where each segment has significantly different content than all the other segments.

After deriving these set of sampled video segment, \Ours independently searches for a set of knob configurations $\mathcal{K}_s$ that lie on an (approximated) work-quality pareto frontier for each segment $s\in\mathcal{S}$. These sets may differ for segments with different content dynamics (e.g., on segments where cheap configurations achieve perfect quality, expensive configurations are not on the pareto frontier). \Ours uses greedy hill climbing~\cite{russell2003norvig} for the search, whose effectiveness on this task has already been demonstrated in VideoStorm~\cite{videostorm}. Finally, filtered set of knob configurations $\mathcal{K}$ is given by the union of all $\mathcal{K}_s$ for $s\in \mathcal{S}$.

\subsection{Filter task placements}

Recall in Section~\ref{sec: overview} that each knob configuration $k$ is associated with a task graph $G_k$, where each node represents the execution of certain user-provided
model (e.g., an object detection model) and each edge specifies the dependency between nodes (e.g., an object tracking model requires the output from the detection model).
Any node can be placed on on-premises or on-demand cloud hardware and the costs in cloud credits (plus bandwidth cost) and runtimes will be different. 
The objective of \emph{placement optimizer} is to find a set of placements $P_k$ for a knob configuration $k$ that is on the cost-runtime Pareto frontier so that \Ours can pick a desirable placement for a knob configuration during online phase.

We first execute the configuration $k$ on the video segments in $\mathcal{S}^*$ on both completely on-premises and completely on-demand cloud hardware. 
The runtime of on-premises processing, the runtime on the cloud, and the input/output sizes of each node in $G_k$ are used as the node features of the placement graph. 
Then, the \emph{placement optimizer} adopts a well-established approach that is shown to be robust and generalizable for placement optimization~\cite{mirhoseini2017device, addanki2019placeto}. Specifically, it takes the node features of $G_k$ as inputs into a Graph Neural Network (GNN)~\cite{scarselli2008graph, welling2016semi} that passes messages between nodes to learn their correlation and output the new node features of $G_k$ after information aggregation. Then, it uses reinforcement learning (RL) to learn the placement strategy of $G_i$. Specifically, each node in $G_i$ is appended with an additional feature: on-premises, on-cloud, or undecided. All nodes are initialized to be ``undecided''. The RL agent iteratively takes $G_i$ as input and optimally make an ``undecided'' node ``on-premises'' or ``on-cloud'', until all nodes are decided.

\smallskip
\textbf{\textit{Simulator for cloud placements:}}
The placement optimizer typically suggests thousands of placements during the search. Executing each of these placements on real hardware would be extremely time- and money-consuming. We therefore use a simulator to estimate the runtime of different placements and use it for training. The simulator is shown to be very accurate and effective for training the placement optimizer. The details of this simulator is presented at Section~\ref{app:eval-simulator}.

\section{Design decisions}
\label{app:design}

In the following, we evaluate design alternatives of \Ours. Subsection~\ref{app:design_simple} provides more details to the discussion on design challenges in Section~\ref{sec: overview}. Subsection~\ref{app:design_dropin} analyzes alternative, drop-in replacements for components of \Ours.

\subsection{From a simplistic system to \Ours}
\label{app:design_simple}

In the following, we evaluate the designs described in the discussion on design challenges in Section~\ref{sec: overview}. The results are simulated on the COVID workload, using the simulator described in Appendix~\ref{app:simulator}. 

The first, simplistic design in Section~\ref{sec: overview} splits time interval $T$ (over which resources should be rationed) into equal-sized splits $t_i$ of a couple of seconds. In our experiment, $T$ is two days long and each $t_i \in T$ is two seconds long. The idealized approach of Section~\ref{sec: overview} then directly predicts the quality $\accbar(k, t_i)$ of each knob configuration $k \in \mathcal{K}$ on each segment $t_i \in T$. In the evaluation in Figure~\ref{fig:app-simple}, we use the average time-of-day quality over the past 2 days as a prediction of the quality over the next 2 days. Fitting a more complicated statistical model (e.g. neural network) is too hard since the output has a dimension of 259.200.


\begin{figure}
    \centering
    \includegraphics[width = 0.46\textwidth]{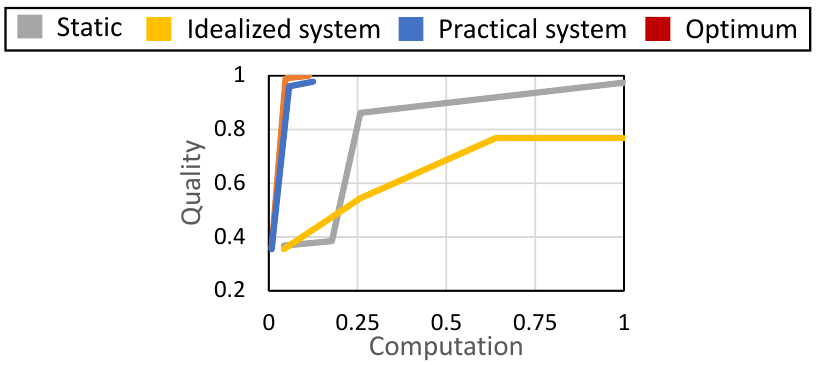}
    \caption{Performance of the systems described in Section~\ref{sec: overview}}
    \label{fig:app-simple}
\end{figure}

Section~\ref{sec: overview} then discusses a design built around a more practical forecasting task. This design resembles \Ours, with the sole difference of \Ours furthermore considering real hardware resources like on-premise compute cores, a video buffer and on-demand cloud workers. Figure~\ref{fig:app-simple} shows this improved design in blue. We can see that the improved design almost achieves optimal quality.

\subsection{Alternative implementations of \Ours components}
\label{app:design_dropin}

In the following, we consider drop-in replacements for specific components of \Ours.

\begin{figure}
    \centering
    \includegraphics[width=0.26\textwidth]{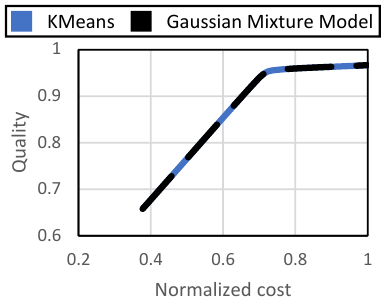}
    \caption{\Ours using different clustering algorithms to compute the content categories.}
    \label{fig:gmm}
\end{figure}

\para{Clustering for content categories}
Figure~\ref{fig:gmm} shows a comparison of the end-to-end performance when using KMeans~\cite{kmeans} for clustering in comparison to using a Gaussians Mixture Model~\cite{gaussianmix}. The experiment is conducted on simulated hardware (simulator described in Appendix~\ref{app:simulator}) and for the COVID workload. We find that there is no difference in the end-to-end performance and therefore suggest using KMeans because it is simpler. 

\para{Forecasting model} \Ours uses a simple feedforward neural network for forecasting. Figure~\ref{fig:eval-forecast} shows that on the COVID and MOT workloads, \Ours's end-to-end performance wouldn't improve even when using a ground truth forecast. We have not seen real-world workloads where the inaccurate predictions by the feedforward neural network cause significant losses of end-to-end performance. We therefore suggest using the feedforward neural network. However, there might exist some workloads where a more sophisticated network architecture might be required.

\para{Computing the knob plan} \Ours assigns knob configurations to content categories using linear programming. Linear programming finds the optimal solution to the optimization problem. Furthermore, Figure~\ref{fig:overheads} shows that the runtime overheads caused be the linear programming solver are negligible: A forward and backward pass through the forecasting model together with solving the linear program take a total of 0.5s or below. This overhead only occurs every couple of days when the knob plan is updated. Therefore, isn't any margin for improvement through alternatives to linear programming.

\section{Numeric values of cost analysis}
\label{app:chameleon}

Figure~\ref{fig:chameleon} in Section~\ref{sec:eval-cost} shows the cost-quality trade off curves of \Ours, Chameleon$^{*}$ (an adaption of Chameleon~\cite{chameleon}) and statically using the same knob configuration. Table~\ref{tab:chameleon} furthermore gives the numeric values of the costs and qualities.

\begin{table}[h]
    \scalebox{0.74}{
    \begin{tabular}{|c|c|c|c|c|c|}
        \thickhline
        \textbf{Workload} & \textbf{Method} & \textbf{Quality} & \textbf{Server vCPUs} & \textbf{Cloud cost} & \textbf{Total cost} \\
        \thickhline
       
        \multirow{11}{*}{COVID} & \multirow{5}{*}{Static} & 35\% & 4 & -- & 14.9\$ \\ \cline{3-6}
        & & 35\% & 8 & -- & 28.8\$ \\ \cline{3-6}
        & & 81\% & 16 & -- & 57.6\$ \\ \cline{3-6}
        & & 81\% & 32 & -- & 114.1\$ \\ \cline{3-6}
        & & 97\% & 60 & -- & 267.7\$ \\ \cline{3-6}
        \cline{2-6}
         & \multirow{4}{*}{Chameleon*} & 37\% & 4 & -- & 14.9\$ \\ \cline{3-6}
        & & 50\% & 8 & -- & 28.8\$ \\ \cline{3-6}
        & & 74\% & 16 & -- & 57.6\$ \\ \cline{3-6}
        & & 91\% & 32 & -- & 114.1\$ \\ \cline{3-6}
        \cline{2-6}
         & \multirow{2}{*}{Skyscraper} & 90\% & 4 & 0.0\$ & 14.9\$ \\ \cline{3-6}
        & & 94\% & 8 & 3.3\$ & 32.1\$ \\ \cline{3-6}
        \thickhline
        
        \multirow{11}{*}{MOT} & \multirow{5}{*}{Static} & 36\% & 4 & -- & 14.9\$ \\ \cline{3-6}
        & & 79\% & 8 & -- & 28.8\$ \\ \cline{3-6}
        & & 81\% & 16 & -- & 57.6\$ \\ \cline{3-6}
        & & 81\% & 32 & -- & 114.1\$ \\ \cline{3-6}
        & & 97\% & 60 & -- & 267.7\$ \\ \cline{3-6}
        \cline{2-6}
        
          & \multirow{4}{*}{Chameleon*} & 72\% & 4 & -- & 14.9\$ \\ \cline{3-6}
        & & 83\% & 8 & -- & 28.8\$ \\ \cline{3-6}
        & & 89\% & 16 & -- & 57.6\$ \\ \cline{3-6}
        & & 92\% & 32 & -- & 114.1\$ \\ \cline{3-6}
        \cline{2-6}
        
          & \multirow{2}{*}{Skyscraper} & 94\% & 4 & 0.0\$ & 14.9\$ \\ \cline{3-6}
        & & 97\% & 8 & 2.0\$ & 30.8\$ \\ \cline{3-6}        
        \thickhline
        
        \multirow{15}{*}{MOSEI-HIGH} & \multirow{5}{*}{Static} & 8\% & 4 & -- & 3.7\$ \\ \cline{3-6}
        & & 8\% & 8 & -- & 7.2\$ \\ \cline{3-6}
        & & 28\% & 16 & -- & 14.4\$ \\ \cline{3-6}
        & & 36\% & 32 & -- & 28.5\$ \\ \cline{3-6}
        & & 51\% & 60 & -- & 66.9\$ \\ \cline{3-6}
        \cline{2-6}
        
         & \multirow{5}{*}{Chameleon*} & 8\% & 4 & -- & 3.7\$ \\ \cline{3-6}
        & & 21\% & 8 & -- & 7.2\$ \\ \cline{3-6}
        & & 32\% & 16 & -- & 14.4\$ \\ \cline{3-6}
        & & 37\% & 32 & -- & 28.5\$ \\ \cline{3-6}
        & & 55\% & 60 & -- & 66.9\$ \\ \cline{3-6}
        \cline{2-6}
        
         & \multirow{5}{*}{Skyscraper} & 30\% & 4 & 0.0\$ & 3.7\$ \\ \cline{3-6}
        & & 38\% & 8 & 0.0\$ & 7.2\$ \\ \cline{3-6}
        & & 45\% & 16 & 0.0\$ & 14.4\$ \\ \cline{3-6}
        & & 59\% & 32 & 0.0\$ & 28.5\$ \\ \cline{3-6}
        & & 80\% & 60 & 0.0\$ & 66.9\$ \\ \cline{3-6}
        \thickhline
        
        \multirow{14}{*}{MOSEI-LONG} & \multirow{5}{*}{Static} & 30\% & 4 & -- & 3.7\$ \\ \cline{3-6}
        & & 30\% & 8 & -- & 7.2\$ \\ \cline{3-6}
        & & 38\% & 16 & -- & 14.4\$ \\ \cline{3-6}
        & & 38\% & 32 & -- & 28.5\$ \\ \cline{3-6}
        & & 65\% & 60 & -- & 66.9\$ \\ \cline{3-6}
        \cline{2-6}
        
         & \multirow{5}{*}{Chameleon*} & 30\% & 4 & -- & 3.7\$ \\ \cline{3-6}
        & & 31\% & 8 & -- & 7.2\$ \\ \cline{3-6}
        & & 39\% & 16 & -- & 14.4\$ \\ \cline{3-6}
        & & 52\% & 32 & -- & 28.5\$ \\ \cline{3-6}
        & & 68\% & 60 & -- & 66.9\$ \\ \cline{3-6}
        \cline{2-6}
        
         & \multirow{4}{*}{Skyscraper} & 37\% & 4 & 1.7\$ & 5.4\$ \\ \cline{3-6}
        & & 53\% & 8 & 3.3\$ & 10.5\$ \\ \cline{3-6}
        & & 62\% & 16 & 6.5\$ & 20.9\$ \\ \cline{3-6}
        & & 72\% & 32 & 12.9\$ & 41.4\$ \\ \cline{3-6}
        \thickhline
    \end{tabular}
    }
    \caption{Cost and quality of running the workloads in Section~\ref{sec:eval-workloads}.}
    \label{tab:chameleon}
    \vspace{-2em}
\end{table}

\section{Ingesting multiple streams}
\label{app:multistream}

While Sections~\ref{sec: overview}, \ref{sec: offline}, \ref{sec: online} focus on how \Ours optimizes the ingestion of a single stream, \Ours's techniques naturally extend to a multi-stream scenario. In this scenario, the overall result quality and the budget of cloud credits is defined across streams, meaning that \Ours must allocate resources between streams to maximize the joint quality. We distinguish between two set ups: First, we analyze the setting where only cloud credits are shared between streams but each stream runs on individually provisioned on-premise compute and has its own buffer. Second, we analyze the setting where multiple streams share an on-premise server (i.e. run on the same cluster) and share a buffer.


For the first scenario, the offline phase is run independently for each stream. Specifically, for each stream, knob configurations and task placements are filtered independently, the content categories are computed independently, and the stream's forecasting model is trained independently of other streams. 
Then, in the online phase, knob switching can also be performed independently for every stream. The only component of \Ours that needs to be modified is the knob planner. This is necessary to ensure that cloud credits are allocated fairly between the streams. Algorithmically, the joint knob planner only presents a slight modification to generalize the knob planner used in the single-stream setting: Let $\mathcal{V}$ be the set of video streams to be ingested. The quality and cost must now be summed over all video streams as shown in Equation~\ref{linpro1-multi} and Equation~\ref{linpro2-multi} respectively. Furthermore, the normalization must occur for all content categories of all streams, as shown in Equation~\ref{linpro3-multi}. We have highlighted the changes compared to the single-stream knob planner in green.

\vspace{-1em}
\begin{alignat}{5}
  & \text{maximize }   & \quad & \mathbin{\textcolor{mygreen}{\sum_{v \in \mathcal{V}}}} \sum_{k,c\textrm{ for }v}\alpha_{ k,c} * r_c * \accbar(k, c)                                                 & \label{linpro1-multi} \\
  & \text{subject to } & \quad & \mathbin{\textcolor{mygreen}{\sum_{v \in \mathcal{V}}}} \sum_{k,c\textrm{ for }v} \alpha_{k,c} * r_c * cost(k) \leq budget                                            & \label{linpro2-multi} \\
  &                    & \quad & \sum_{ k} \alpha_{k,c} = 1,    \ \ \ \alpha_{k,c} \geq 0      \quad \hspace{6pt}                \forall c \textcolor{mygreen}{\textrm{ of all } v \in \mathcal{V}}  & \label{linpro3-multi}
\end{alignat}
\vspace{-1em}

If multiple streams are processed with shared on-premise resources (i.e. the second scenario), it is sufficient to use the previous approach with two slight modifications. First, since multiple streams now share a buffer, each independent knob switcher must be aware of the true free capacity of the buffer, even if some memory is occupied by other streams. Therefore, the knob switcher implementation cannot assume anymore that it is the only one that allocates and deallocates memory in the buffer. Second, for the (offline) placement optimization, it is unclear how many cores are allocated to each stream. We propose a simple solution to this: For $n$ cores and $|\mathcal{V}|$ streams, we can simply assume that each stream is (fairly) allocated $\lfloor n\, / \, |\mathcal{V}| \rfloor$ cores. 

Note that assuming a fair allocation of the cores between streams precludes buffer overflows while not leading to under-utilization of the on-premise cores: The fair allocation is a pessimistic estimate because it assumes that every stream's workload is high at the same time (i.e. every stream can keep all $\lfloor n\, / \, |\mathcal{V}| \rfloor$ cores busy). Since \Ours has the overall assumed $n$ cores available for processing, progress won't be slower than estimated which allows \Ours to avoid buffer overflows. 

Nevertheless, the fair allocation assumption doesn't lead to under-utilization when a stream does not utilize its $\lfloor n\, / \, |\mathcal{V}| \rfloor$ cores. Since the mapping of tasks to cores is independent of the fair allocation assumption (e.g. Ray performs the mapping in our \Ours implementation), cores that are unused by one stream can still be used to run tasks of another stream. 
In that case, \Ours will overestimate the runtime of the knob configuration since work can be shared among more cores than assumed. However, this doesn't matter because the knob switcher is reactive and will notice that the buffer fills slower than expected. Continuing to have available buffer space, the knob switcher will forgo placing tasks on the cloud and the overestimated runtimes will therefore not cause unnecessary cloud spending. All of this doesn't require any additional changes to the knob switcher.




\section{Runtime of offline components}
\label{app:offline_scale}

\subsection{Runtime of offline components}

In the following, we evaluate the cost of the offline phase for the COVID workload. The runtime is dominated by processing the training data with the user-defined processing DAGs of the different knob configurations. Table~\ref{tab:offline_runtimes} shows how much time was spent for each step in the offline phase in the experiments conducted in Section~\ref{sec:eval}. The runtimes are measured when using two c2-standard-60 Google Cloud instances for the offline phase, which cost a total of 7.8\$. Note that creating the training data for the forecasting model is embarrassingly parallel, therefore more machines will result in lower runtimes.

\begin{table}[h]
    \centering
    \begin{tabular}{l l}
    \toprule
    \textbf{Step} & \textbf{Runtime} \\
    \toprule
        Filter knob configurations & 6 min \\
        Filter task placements & 4 min \\
    \midrule
        Compute content categories & 5 min \\
    \midrule
        Create forecast training data & 1.3 h \\
        Train forecast model & 1 min \\
    \bottomrule
    \end{tabular}
    \caption{Runtimes of the offline steps for the COVID experiments conducted in Section~\ref{sec:eval}}
    \label{tab:offline_runtimes}
\end{table}






In Table~\ref{tab:offline_runtimes} and in our experiments in Section~\ref{sec:eval}, we processed 16 days of video data with the cheapest knob configuration $k^{-}$ to create 1200 training samples in 1.3 hours (as described in Section~\ref{sec: offline}). However, the forecasting model can also be trained with less data, which requires less video to be recorded up front and reduces the work required to create the training data. In Figure~\ref{fig:forecast_samples}, we show the Mean Absolute Error (MAE) when training the forecasting model with varying amount of training samples. As shown in Figure~\ref{fig:forecast_samples}, training the forecasting model with only 700 samples wouldn't have increased the model's MAE but would have reduced the runtime of the offline phase by 35\%.

\begin{figure}[h]
    \centering
    \includegraphics[width=0.28\textwidth]{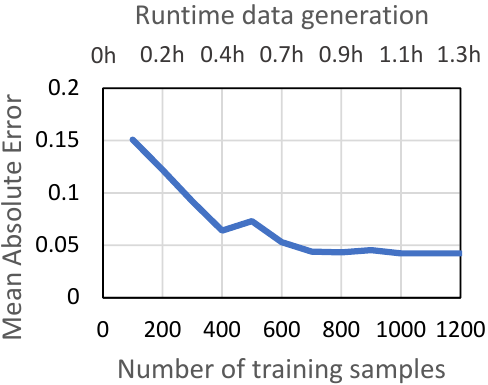}
    \caption{Mean Absolute Error of the forecasting model for different amounts of training data}
    \label{fig:forecast_samples}
    \vspace{-1em}
\end{figure}

\subsection{Rerunning the offline phase}

Generally the offline phase only needs to be run once. However, under rare circumstances some of the components might need to be rerun online. While these circumstances should never occur for most workloads, we discuss in the following when which offline component would need to be rerun.

\para{Retraining the forecasting model} If \Ours is set up properly, training the forecasting model is the only step of the offline phase that might need to be redone. How often different content categories appear may change over time (e.g. traffic in the city worsens). These changes are mostly smooth and should not be problematic since we continuously train the forecasting model online ("online learning"). This continuous training allows the model to adapt to smooth drifts at negligible overhead. \Ours can monitor the forecasting accuracy online (the knob switcher determines the actual frequency of content categories anyways). If this accuracy deteriorates and the user decides to retrain the model, this is cheap since the costly part of generating the new training data has already been done anyways during video ingestion.

\para{Content categories} Typically, workload changes only involve how frequently different content categories appear, but don’t introduce completely new content categories (e.g. there is no "completely new type of heavy traffic"). However, if the training data is incomplete and does not contain a content category, the content categories might need to be recomputed. \Ours can detect this, as the measured quality will then frequently be far from all of the KMeans cluster centers. Recomputing the content categories is cheaper than in the original offline phase since the sample efficiency can be drastically improved. Instead of randomly sampling segments and computing their quality vectors, \Ours can simply add segments to the train set, where it notices that the samples don't belong to any of the content categories.

\para{Filtered set of good knob configurations \& good task placements} Task placements are independent of the video and don’t need to be recomputed. The user may consider (partially) recomputing the set of knob configurations if the content categories have been recomputed. As described above, this should generally not happen.

\section{\Ours API}
\label{app:api}

To support application-specific query formats, \Ours lets users define processing steps as arbitrary user-defined functions (UDFs). When applying the UDFs to the video stream, one UDF typically produces the input for another, yielding a directed-acyclic graph (DAG) where each node is a UDF and edges specify that the source UDF's output is used as the target UDF's input. The code snippet below shows the relevant lines of code of an implementation of the EV counting example of the Introduction.


\begin{lstlisting}
# Skyscraper app (UDFs yolo, yolo_cloud, kcf, kcf_cloud omitted)
def proc_frame(frame, frame_num, sky, state):
  # get knobs
  det_interval = sky.knob("det_interval")
  yolo_size = sky.knob("yolo_size")
  # process depending on knob values
  if frame_num % det_interval.val() == 0:
    state = sky.run(yolo, yolo_cloud, frame, yolo_size)
  else:
    state = sky.run(kcf, kcf_cloud, frame, state)
  return state

# instantiate Skyscraper
sky = Skyscraper(aws_key_id, aws_secret_key, fps=30)
sky.set_resources(num_cores=8, bufferMB=4000, cloud_budget=1000)

# register knobs
sky.register_knob("det_interval", [1, 5, 10])
sky.register_knob("yolo_size", ["small", "medium", "large"])

# offline preparation
sky.fit(labeled_video, labels, unlabeled_video, proc_frame)

# online ingestion (e.g., cv2 to read frames)
state = State(init_quality=0)
vid = cv2.VideoCapture(0)
ok, frame = vid.read()
while ok:
  status, state = sky.process(frame, state)
  ok, frame = vid.read()
\end{lstlisting}

Lines~14-15 of the code snippet instantiate the \verb|Skyscraper| instance \verb|sky| to process a video stream.
The user registers application-specific knobs to \verb|sky| by specifying the knob's name and a value domain that the knob can take. Specifically, the user registers a knob that determines the rate at which a object detector is run (line~17) and a knob that determines its model size (line~18).

The application only uses two UDFs, namely \verb|yolo| and \verb|kcf|. For brevity, we omit their implementation but users would typically implement them using popular CV libraries (e.g. \verb|torchvision|, \verb|OpenCV|). The user needs to specify an on-premise version and a cloud version for each UDF. \Ours then calls the corresponding version depending on if it wants to execute the UDF on premises or in the cloud. Currently, the user is responsible for packetizing the payload and invoking the cloud function within the cloud UDF.

The user defines the processing DAG applied to the video in \verb|proc_frame|. The \verb|state| object is used to carry state between frames (in this case, it is used to keep track of bounding boxes that both \verb|yolo| and\verb|kcf| update). The \verb|state| object is user-defined but must contain a \verb|quality| field which the user updates to reflect the current quality of processing.  Quality is an application-defined metric and may for example be based on errors or certainty metrics that are commonly reported by CV algorithms. Specific examples for quality defintions can be found in Section~\ref{sec:eval-workloads}, where we describe the \Ours applications that we use for benchmarking. 

The user triggers \Ours's offline learning phase in line~22.
After its completion, the user can start to ingest live videos as shown in lines~28-30.

\section{Comparison to query-load- adaptive knob tuning systems}
\label{app:videostorm}

VideoStorm~\cite{videostorm} and VideoEdge~\cite{videoedge} perform knob tuning for video workloads but only adapt to the query load and not the content of the video. We discuss this in more detail in Section~\ref{sec:introduction} and Section~\ref{sec:rel}. Since VideoEdge is designed for a different compute hierarchy, we only compare to VideoStorm in the following. We want to emphasize that the following results are measured on the \vetl workloads described in Section~\ref{sec:eval}, which significantly differ from the workloads that VideoStorm was designed for. The following results are not representative of VideoStorm's performance on workloads it was designed for (i.e. many finite, ad-hoc queries being run on a video stream). Instead, they provide experimental evidence that systems that only adapt to the query load are not suitable for \vetl.

\begin{figure}
    \centering
    \includegraphics[width=0.33\textwidth]{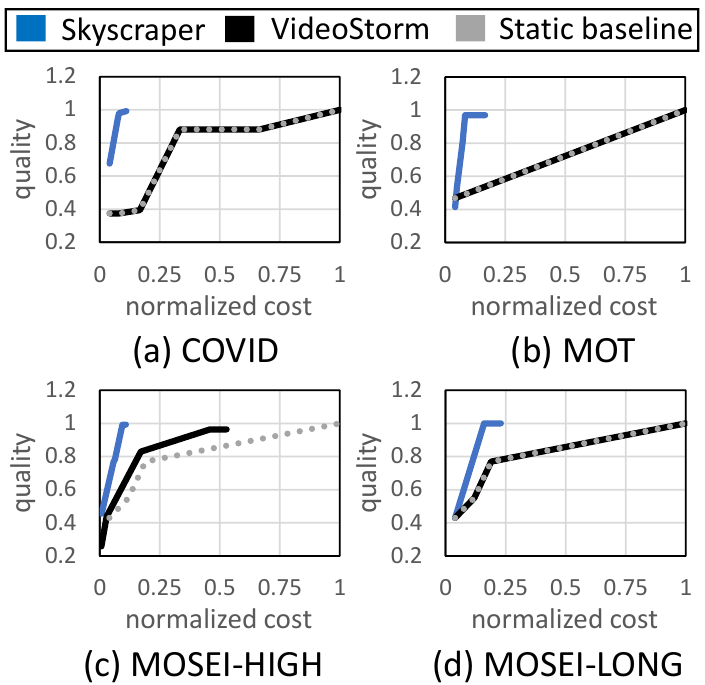}
    \caption{Comparison between \Ours, the static baseline and VideoStorm. VideoStorm was designed for significantly different workloads.}
    \label{fig:videstorm}
    \vspace{-1em}
\end{figure}

Figure~\ref{fig:videstorm} shows the cost-quality trade offs when statically using the same knob configuration, VideoStorm and \Ours. For all workloads, VideoStorm fills the buffer early during execution and then (almost) only uses the most qualitative knob configuration that runs in real-time on the provisioned hardware. VideoStorm's performance therefore closely matches the one of the static baseline.

The reason why VideoStorm outperformed the static baseline for MOSEI-HIGH, is that the first workload spike in MOSEI-HIGH comes early enough, such that VideoStorm hadn't used up significant buffer space yet. By being able to leverage the buffer during the workload peak, MOSEI-HIGH could therefore achieve better quality on cheaper resources than the static baseline. However, it is a lucky coincidence that VideoStorm hadn't used up the buffer at the beginning of the workload peak and this performance improvement can only be observed for the first peak. For all subsequent data, VideoStorm will have the performance of the static baseline since the buffer remains full after the first peak.

\section{Details on training the forecasting model}
\label{app:train-data}

\Ours pre-trains the forecasting model in the offline phase using the unlabeled data. In the following, we describe how \Ours computes the training data from the unlabeled data. Previously, \Ours has already computed the content categories $\mathcal{C}$. 

First, \Ours processes all of the unlabeled training data using the cheapest knob configuration $k^-$.\footnote{If $k^-$ achieves similar performance for different content categories (i.e. is not a good discriminator), the next cheapest configuration is picked that is a good discriminator.} Using the quality that $k^-$ achieves on each of the segments of the unlabeled training data, \Ours classifies the segments into one of the content categories $c\in\mathcal{C}$. This is done through \Ours's standard way of classifying content, which is described in Section~\ref{sec:knob-switch}.

Given the category of each segment, \Ours can create inpu-label pairs which are used to train $\mathcal{F}$ via supervised learning. Each input $x$ spans a time period of $t_{in}$ into the past and each label spans a time period of $t_{out}$ into the future. The label is given by one content histogram that contains the content distribution over the $t_{out}$ long interval. The input contains $n_{split}$ histograms that contain the content distributions of $t_{in}/n_{split}$ time chunks that the past $t_{in}$ interval has been split up to.

$t_{out}$ (planned interval length), $t_{in}$ (forecast model input length) and $n_{split}$ (number of forecast model input splits) are hyperparameters of \Ours but as discussed in Appendix~\ref{app:hyperparams}, setting them to default values generally leads to good performance, so we envision that the users do not need to tune them.

\section{\Ours's hyperparameters}
\label{app:hyperparams}

\Ours exposes the following set of hyperparameters that can be adjusted by the user:

\begin{enumerate}
    \item Number of content categories 
    \item Frequency of knob switching
    \item Input features for forecasting model:
    \begin{enumerate}
        \item Input time span for forecasting model ($t_{in}$)
        \item Number of histograms reported for the input ($n_{split}$)
    \end{enumerate}
    \item Hyperparameters of the forecasting model (e.g. architecture, training hyperparameters)
    \item Planned interval length ($t_{out}$)
    \item Sample sizes in the offline phase
    \begin{enumerate}
        \item Sample size of $\mathcal{S}$ for searching good knob configurations $\mathcal{K}$
        \item Sample size of $\mathcal{S'}$ for computing content categories $\mathcal{C}$
    \end{enumerate}
\end{enumerate}

In the following, we describe how to set these hyperparameters to ensure \Ours achieves a good performance. We hereby suggest default values that worked well on all four workloads considered in the paper. Our tuning recommendations are supported by sensitivity analyses conducted in the subsections below.

\begin{enumerate}
    \item \textbf{Number of content categories $\mathcal{C}$:} This parameter determines how many clsuter centers should be used in KMeans. \Ours's performance is insensitive to this as long as it is set high enough. For our workloads, values of 3 and above performed well. We evaluate different values in Subsection~\ref{app:sens-kmeans} We suggest a default value of 4.
    \item \textbf{Frequency of knob switching:} While \Ours's performance is sensitive to this hyperparameter, we find that reasonable values (running it between every 2s to 8s) all achieve good performance. We evaluate the performance for different values in Subsection~\ref{app:sens-switchfreq}. We suggest a default of running it every 4s.
    \item \textbf{Input features for forecasting model:} We find that the most important property of the featurization is that the model knows about the content dynamics of the recent past (and these dynamics have not been averaged over a long time period). In Subsection~\ref{app:sens-forecast}, we find that any featurization that fulfills this delivers results that are accurate enough to not harm \Ours's performance. We suggest a default of providing data from the previous two days as input, split into eight histograms.
    \item \textbf{Hyperparameters of the forecasting model} We find that very simple architectures and training procedures are sufficient. We suggest to use the same as we used in our experiments as default (see Appendix~\ref{app:eval-detail}).
    \item \textbf{Planned interval length}: We evaluate this in Section~\ref{sec:eval} and find that \Ours performs well for reasonable values between 1 day to 4 days. We suggest a default of 2 days.
    \item \textbf{Sample sizes in offline phase:} For both the sample size of $\mathcal{S}$ (filter knob configurations) and $\mathcal{S'}$ (categorize video content) of the offline phase, larger sample sizes are better but cause a longer runtime of the offline phase. We suggest a sample size of 5 for  $\mathcal{S}$ and a sample size of 5\% of the unlabeled training data for  $\mathcal{S'}$.
\end{enumerate}

If users want to tune hyperparameters beyond the default suggestions, \Ours allows for easy hyperparameter tuning in the offline phase. Tuning the hyperparameters hereby only requires rerunning the affected component (e.g. re-training the forecasting model while not re-running anything else). 

\subsection{Sensitivity to the number of content categories}
\label{app:sens-kmeans}

\Ours categorizes content into $|\mathcal{C}|$ content categories. Each category correponds to a cluster center computed through KMeans (the number of content categories therefore corresponds to the "k in KMeans"). Figure~\ref{fig:app-kmeans} shows \Ours's end-to-end performance on the COVID workload for different numbers of content categories.

\begin{figure}[h]
    \centering
    \includegraphics[width=0.8\linewidth]{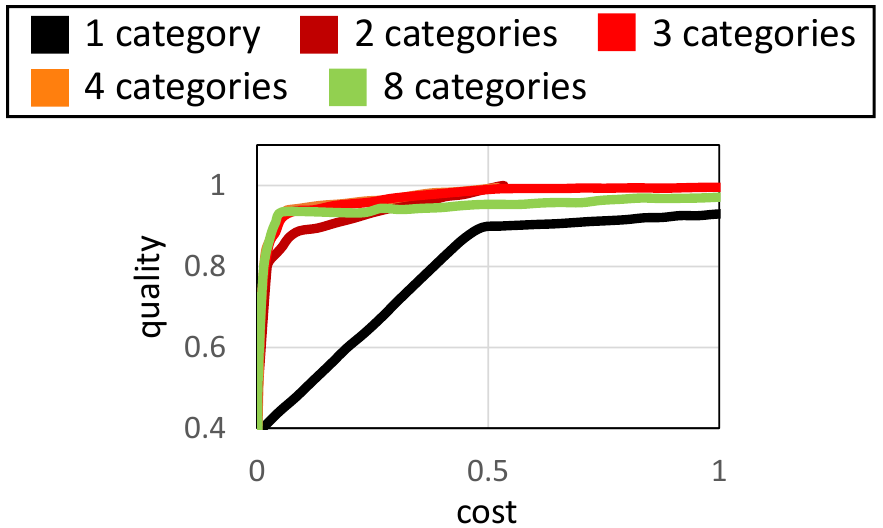}
    \caption{\Ours's end-to-end performance on the COVID workload using a varying number of content categories.}
    \label{fig:app-kmeans}
\end{figure}

Figure~\ref{fig:app-kmeans} shows how \Ours's performance is insensitive to the number of content categories as long as it is high enough (assuming no degenerately high numbers). Table~\ref{tab:app-switcher} shows the classification accuracy of the knob switcher for different numbers of content categories.
\begin{table}[h]
    \centering
    \begin{tabular}{|c|c|}
    \hline
    \textbf{Method} & \textbf{Switcher accuracy} \\
    \hline
       1 category  &  100\% \\
    \hline
       2 categories  & 98.8\% \\
    \hline
       3 categories  & 97.9\% \\
    \hline
       4 categories  & 97.2\% \\
    \hline
       8 categories  & 95.9\% \\
    \hline
    \end{tabular}
    \caption{The accuracy of the knob switcher on the COVID workload for a varying numbers of content categories}
    \label{tab:app-switcher}
\end{table}

\subsection{Sensitivity to the knob switching frequency}
\label{app:sens-switchfreq}

\Ours periodically runs the knob switcher every couple of seconds. Figure~\ref{fig:app-switcher} shows \Ours's end-to-end performance of \Ours when running the knob switcher at different periodicities. We can see that \Ours is sensitive to the periodicity but the performance variance between different periodicities is not high.

\begin{figure}[h]
    \centering
    \includegraphics[width=0.8\linewidth]{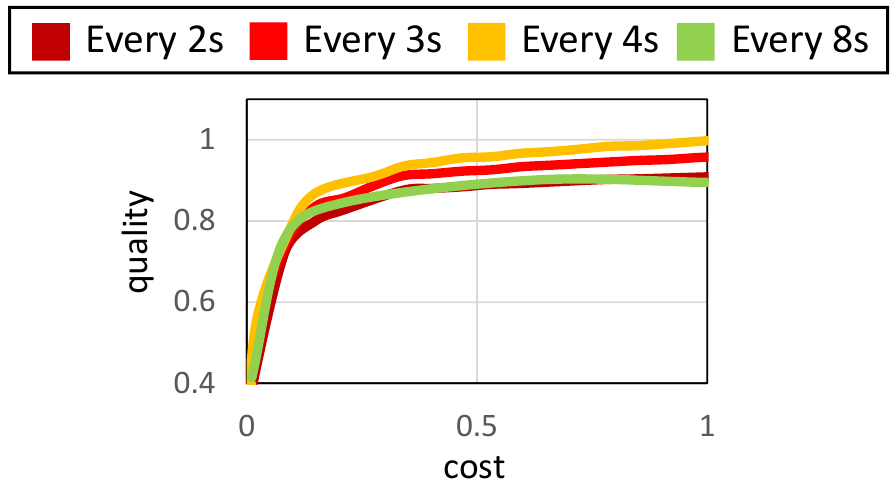}
    \caption{The effect of running the knob switcher at different frequencies on \Ours's end-to-end performance}
    \label{fig:app-switcher}
\end{figure}

\subsection{Sensitivity of forecasting model }
\label{app:sens-forecast}

Table~\ref{tab:app-mae-output} complements Figure~\ref{fig:eval-forecast} and shows the Mean Absolute Error (MAE) of forecasts over different time intervals. As shown in Figure~\ref{fig:eval-forecast}, only the prediction error for 8 days caused significantly harmed performance.

\begin{table}[h]
    \centering
    \begin{tabular}{|c|c|c|}
    \hline
    
    \multirow{2}{*}{\textbf{Days forecasted}} & \textbf{COVID Mean} & \textbf{MOT Mean} \\
             & \textbf{Absolute Error} & \textbf{Absolute Error} \\
    \hline
       1 day  & 0.097
 & 0.108
 \\
    \hline
       2 days  & 0.042
 & 0.064
 \\
    \hline
       4 days  & 0.066
 & 0.133
 \\
    \hline
       8 days  & 0.149
 & 0.185
 \\
    \hline
    \end{tabular}
    \caption{Mean Absolute Error for varying number of forecasted days, as evaluated in Section~\ref{sec:microbenchmarks}}
    \label{tab:app-mae-output}
\end{table}

Table~\ref{tab:app-mae-input} shows the MAE when predicting over 2 days but with varying amounts of input days (how many days of data are fed into the forecasting model) and splitting the input data into different amounts of content distribution histograms. We can see that if we split the input data into 8 histograms, the MAE is always significantly below what would cause performance deterioration. 

\begin{table}[h]
  \includegraphics[width=0.68\linewidth]{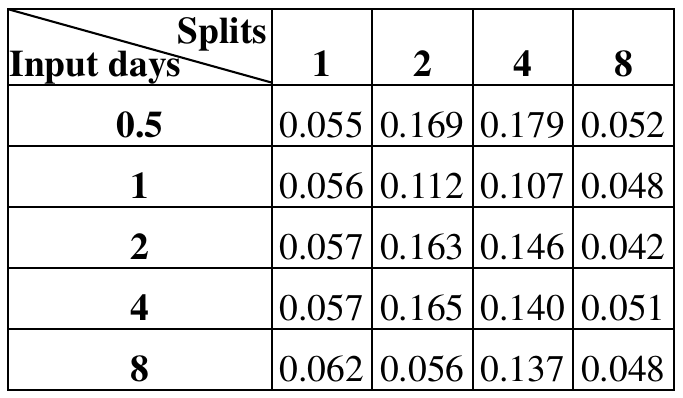}
  \caption{Mean Absolute Error depending on the input features}
  \label{tab:app-mae-input}
\end{table}

\section{Details of the Workloads}
\label{sec:app-workloads}

We evaluate \Ours using three workloads on public health monitoring, traffic planning, and social media analysis. They cover a diverse set of computer vision primitives including object detectors, trackers, and classifiers. All workloads are heavily CPU bound. We describe them in the following.

\para{COVID-19 safety measures (COVID)} During the coronavirus pandemic, decision makers have executed several safety measures to slowdown the spread of the virus. Such measures include wearing facial masks and social distancing. Measuring where and how strictly people adhere to these measures can be used for decision making and informing people at risk. 

The COVID workload consists of a YOLOv5 object detector~\cite{yolo} to detect pedestrians and a KCF tracker to track the detected pedestrians ("detect-to-track"). After the detection, for each detected pedestrian, the workload employs homography to measure the pedestrian's distance to others. Furthermore, it uses a neural network with a ResNet-50~\cite{he2016deep} backbone to classify if the person is wearing a facial mask. The neural network classifier was fine tuned on MaskedFace-Net \cite{MaskedFaceNet}.

The workload contains the following knobs:

$\bullet$ \textit{Frame rate:}
    Consecutive video frames contain large redundancy. It is a common approach to skip frames at a constant rate to effectively reduce this redundancy. We expose the following frame rate domain to \Ours: \{30FPS, 15FPS, 10FPS, 5FPS, 1FPS\}.

$\bullet$ \textit{Object detection rate:}
    In the detect-to-track framework, an object detector is run on frames with a regular interval and cheaper trackers are run on the intermediary frames to track the detected objects. Running object detection more frequently will lead to a better quality and a higher cost. \Ours considers running object detectors every \{1, 5, 30, 60\} frames.
    
$\bullet$ \textit{Tiling for object detection:} The pretrained object detectors are typically trained on small input images to reduce cost (e.g. Ultralytics' YOLOv5 weights). Higher resolution images during video ingestion thus need to be down sampled to the fixed, small dimension used for testing. This impedes the model's ability to detect small objects. A common approach to overcome this issue is to slice the high resolution image into several \textit{tiles}~\cite{yolo-tile} and individually fed them to the object detector. More tiles will thus result in better accuracy but at a higher cost. We expose the following tiling domain to \Ours: \{1x1 tile, 2x2 tiles\}.

The workload is executed on a video stream of a busy shopping street in Tokyo.\footnote{The Koen-Dori street in the Shibuya district. Live stream available at  \url{https://youtu.be/gALQR-nsEME} (July 7, 2022)} We measure the workload's quality in the $person*seconds$ that it records. This metric encompasses that for different knob configurations, people may be detected at different times and different sets of people may be detected. It further leverages that KCF trackers report when they fail to track an object.

\para{Multi-object tracking (MOT)} Multi-object tracking (MOT) is a key primitive in many video analytical pipelines. In this workload, we adopt the recent state-of-the-art TransMOT~\cite{transmot} tracker on MOT benchmark~\cite{MOT20} and introduce several tunable knobs.
TransMOT first runs a object detector and use off-the-shelf image models (such as VGG~\cite{simonyan2014very}) to create feature embeddings of the detected objects. 
TransMOT then models all object features and interactions on one frame as a graph and inputs this graph and the graphs from previous frames to a graph transformer to generate the object tracks.

Apart from the \textit{frame rate} and \textit{number of tiles} explained in COVID workload, MOT contains two additional knobs:

$\bullet$ \textit{Length of history:} TransMOT model takes the graph from previous $t$ frames as history and we set $t$ as tunable knob. Larger $t$ suggests a better quality and a higher cost.

$\bullet$ \textit{Model size:} We trained three TransMOT models with different number of layers (different model size). We can adaptive use any model according to their accuracy and cost trade-offs.

The quality $Q_{MOT}$ is defined as the number of people that TransMOT correctly tracked. The ground truth tracking is given by running TransMOT at the most expensive knob setting, that we do not consider in the experiments. 
We run MOT on a stream of a traffic intersection Shibuya in Tokyo and track pedestrians. 

\para{Multi-modal opinion sentiment and emotion 
intensity (MOSEI)} 
This workload is synthetic and simulates a video stream analysis on Twitch. The number of incoming streams varies over time and mimics the number of live Twitch streams over two days.\footnote{As recorded by Twitch Tracker at \url{https://twitchtracker.com/statistics/active-streamers} (7 July, 2022)} We further introduce synthetic spikes to evaluate \Ours under difficult conditions. Specifically, we create the following two type of spikes:

$\bullet$ \textit{MOSEI-HIGH:} We introduce high but short peaks in workload, consist of 62 concurrent video stream ingestion. This makes cloud bursting difficult since they will require high bandwidth.

$\bullet$ \textit{MOSEI-LONG:} We introduce a long peak of continuous workload. In this case, the buffer alone cannot hold all the extra work.

Since the real data from twitch is not available, we use the CMU-MOSEI~\cite{cmu-mosei-2018-multimodal} dataset to simulate incoming video streams. It contains various videos filming people's heads while talking from YouTube. The task of MOSEI workload is to classify the opinion sentiment of the speaker using both the audio and visual content. CMU-MOSEI provides extracted features from the video with ground-truth labels. We trained a neural network on CMU-MOSEI's training set and used its test set to evaluate \Ours. We couldn't find out which exact methods CMU-MOSEI used for feature extraction but replicated a similar pipeline, which we run before the neural network classifier. We transcribe the audio using CMUSphinx~\cite{cmusphinx} and use GloVe word embeddings~\cite{pennington-etal-2014-glove} on the transcript. 
For the visual features, we extract the bounding box of the face using MTCNN~\cite{mtcnn-zhang} and the face embeddings using DeepFace~\cite{deepface-2014}. We further extract acoustic features including 12 Mel-frequency cepstral coefficients, voiced/unvoiced segmenting features~\cite{voice-segment-drugman-abeer}, and glottal source parameters~\cite{glottal-drugman, glottal-alku-1997, glottal-alku-2002}.
Since CMU-MOSEI does neither provide raw video nor the exact processing steps to obtain their features, we can only simulate the feature extraction pipeline as described in subsection \ref{sec:eval-workloads} but then predict on the features as provided by the CMU-MOSEI dataset. 

The workload contains the following knobs:

$\bullet$ \textit{Frequency of sentiment analysis:} How frequently is sentiment analysis performed. Since the precise sentiment is  volatile, frequent sentiment analysis improves accuracy but is more expensive. Since the workload always transcribes the spoken audio, we determine the frequency based on sentences that are skipped. \Ours may skip \{0, 1, 2, 3, 4, 5, 6\} sentences.

$\bullet$ \textit{Frame rate during sentiment analysis:} What fraction of data is analyzed for each sentence that is chosen for analysis. Given a sentence on which sentiment analysis should be performed, we sample the video frames and the corresponding audio and transcription at regular intervals. This reflects that only analyzing part of a sentence may already reveal its sentiment. We expose the following domain to \Ours: $\{\frac{1}{6}, \, \frac{1}{3}, \, \frac{1}{2}, \, \frac{2}{3}, \, \frac{5}{6}, \, 1\}$
    
$\bullet$ \textit{Model size:} We trained three models of different sizes for the sentiment analysis. The models show a correlation between accuracy and runtime such that slower models have a higher accuracy.

$\bullet$ \textit{Number of streams:} The number of streams to analyze.

When processing $n$ streams and stream $i$ is processed with resulting accuracy $a_i$, we define the quality as $Q_{MOSEI} = \sum_{i=1}^{n}a_i$.

\section{Evaluation details}
\label{app:eval-detail}

In the following, we provide further details on the experiments run in Section~\ref{sec:eval}.

\subsection{Hyperparamters used in evaluation}

In the evaluation in Section~\ref{sec:eval}, all workloads were executed with the same hyperparameter setting except for when otherwise noted.

\begin{enumerate}
    \item Number of content categories: COVID and MOT use 3 content categories, MOSEI-HIGH and MOSEI-LONG use 5.
    \item Frequency of knob switching: For COVID and MOT, the knob switcher is run every 2 seconds, for MOSEI-HIGH and MOSEI-LONG, it is run every 7 seconds (due to constraints of the data set).
    \item Input features for the forecasting model: For all workloads, the forecasting model used 2 days of data split into 8 histrograms as input.
    \item Hyperparameters of forecasting model: All workloads used the following feed-forward architecture: 

    \verb|input --> 16 units (RELU) --> 8 units|
    
    \verb|(RELU) --> num content categs (softmax)| 

    For all workloads, the model was trained for 40 epochs and the weights with the best validation accuracy were chosen for the online phase. The validation split was 20\% for all workloads. COVID and MOT used 16 days of data from a traffic camera in Tokyo\footnote{Live camera stream available at:\url{https://www.youtube.com/watch?v=IKbbHFMAeBM} (accessed on 29 Nov 2022)} as training data and MOSEI-HIGH and MOSEI-LONG used 10 days of synthetically generated data as training data (as described in Section~\ref{sec:eval-workloads}).
    \item For all workloads, the planned interval length was 2 days.
    \item For COVID and MOT, we sampled 4 segments for filtering knob configurations and sampled 5\% of the unlabeled training data to find the content categorization. For MOSEI-HIGH and MOSEI-LONG we sampled 10 segments for the knob configuration filtering and sampled 10\% for the content categorization.
\end{enumerate}

For all workloads, we create a training point for the forecasting model every 15 minutes of data.

\subsection{Decode cost}

In our experiments, we use the default OpenCV decode function and decode frames from H.264 as soon as they enter the system. Using 4 cores on a Intel(R) Xeon(R) Platinum 8260 CPU @ 2.40GHz, we measure that decoding a frame takes 1.6ms, which amounts to taking 5\% of the total runtime for processing. This doesn't play a big role given that the CV models used for inference in our experiments often take around 100ms to run. For example, YOLOv5 on the same hardware configuration takes 86ms per inference.

\section{Cloud vs On-Premise Total Cost of Ownership}
\label{app:cloudcost}

The same computation on the Cloud is generally more expensive than on premise. This section estimates the cost ratio between Cloud and on premise compute. We hereby follow Greg Deckler’s estimate but use current prices and simplify further.\footnote{\url{ https://www.linkedin.com/pulse/cloud-vs-on-premises-hard-dollar-costs-greg-deckler/} (accessed on 14 July 2022)} The simplifications are to the disadvantage of the cloud and make the cloud compute more expensive when compared to the on premise compute. This is to the disadvantage of Skyscraper, as Skyscraper's cloud costs are multiplied by a higher factor.  We take the following simplifying assumptions:

\begin{itemize}
    \item Setting the on premise hardware up and maintaining it is for free. There are no staff costs and also no damages or other maintenance costs.
    \item We ignore tax implications (cap-ex vs. op-ex)
    \item We assume a 3 year (36 months) lifecycle of all hardware. This matches Greg Deckler’s assumption and is common in similar cost analyses.
    \item We ignore software licensing costs for on premise.
    \item A month has 744 hours.
    \item We ignore cost for space (i.e. rent) for on premise.
    \item We ignore costs for networking hardware (e.g. network switches).
\end{itemize}

We compare the on-premise cost to the AWS Lambda 3000MB instance that we used in our experiments. Renting one of these instances over an entire month costs 130.78 USD/month.\footnote{\url{ https://aws.amazon.com/de/lambda/pricing/} (accessed on 14 July 2022)}
For the on-premise hardware, we consider the Dell R240 as a cheap commodity sever at the time of writing. In the cheapest, default configuration, Dell states the value of this machine at 1596.90 USD.\footnote{\url{https://www.dell.com/en-us/work/shop/cty/pdp/spd/poweredge-r240/pe_r240_13157_vi_vp} (accessed on 14 July 2022)} This configuration includes an Intel Celeron G4930 with 2 cores. Since computer vision tasks are generally compute bound and parallelizable, the number of cores largely determines the runtime. We found that our AWS Lambda instance uses 2 Intel Xeon cores and is therefore comparable. Dividing the cost of the Dell R240 server by its assumed life cycle of 36 months gives a monthly cost of 47.2 USD.
In the standard configuration, the Dell R240 comes with a 250W power supply. In April 2022, the average electricity cost across all US states was 15.38 cents per kWh.\footnote{\url{ https://www.eia.gov/electricity/monthly/epm_table_grapher.php?t=epmt_5_6_a} (accessed on 14 July 2022)} Using 250W over a month of 744 hours results in 186 kWh, which cost 28.6 USD/month.
In total, this gives a Cloud to on premise cost ratio of $1.8\times$.

\section{Simulator}
\label{app:simulator}

In the following, we describe the simulator algorithm in detail and evaluate it.

\subsection{Simulator algorithm}
\label{app:sim-desc}

The placement optimizer typically suggests thousands of placements during the search. Executing each of these placements on real hardware would take a long time and require the user to pay money to execute the tasks that the optimizer placed on the cloud. 
To make the placement search practical, \Ours instead uses a simulator to estimate the runtimes of a given placement.

The simulator hereby takes a directed acyclic task graph (DAG) as input where each node is a UDF that is labeled for execution on cloud or on premises.

Before simulating placements of the DAG, the simulator profiles each UDF on real hardware. The simulator hereby measures the following three properties:

\begin{itemize}
\item \textbf{Runtime on 1 on premise core:} Some UDFs are multi-threaded and can run on several cores in parallel. During video ingestion however, other cores are typically occupied by other UDFs and a UDF may hence often only utilize one core. Because of this, the simulator assumes that each UDF is scheduled on a single core. To obtain the runtime of a UDF on a machine with $n$ cores, we measure the runtime of executing $n$ UDF instances on the machine in parallel. Like this, each UDF is usually scheduled to run on one core and we measure the runtime as desired.

\item \textbf{Round trip time for cloud version of UDF:} As described in section 2, each UDF also has a version where most of the processing is done on the cloud. Measuring the runtime of this function is trivial as it is largely dominated by the round trip time to the cloud, which includes the processing time on the cloud. We assume that the function does not experience a cold start and therefore warm up cloud workers before measuring.

\item \textbf{Average input \& output sizes of UDF in bytes:} To estimate the bandwidth requirement of tasks, we measure the average size of the payload that the function uploads to the cloud and receives from the cloud. In video processing, we observe that the tasks' input and output sizes do not vary much.
\end{itemize}

The input to a simulation then consists of two things: First, the simulator takes a directed acyclic task graph (DAG) as input, where the nodes are measured tasks and the edges are inter-task dependencies. Second, the simulator takes a placement of the task graph where each task in the DAG is either placed on premises or on the cloud. The simulator then outputs the runtime of this placement. 

The estimated runtime is the time $t_{max}$ at which the simulator estimates the last task to finish. The simulator iteratively simulates the execution of tasks and updates $t_{max}$ accordingly. Initially, $t_{max} = 0$. In each iteration, the simulator then chooses the task $T$ for execution, whose dependencies are resolved at the earliest time $t^{ready}_T$. In the first iteration, this is a task with no dependencies which exists by the definition of DAG. The simulator keeps track of when the last cloud task finishes through $t_{max}^{cloud}$ and also keeps track for each on premise core $c$, at what time $t_{max}^c$ the last task on that core finishes.

For a task $T$ that is placed on premise, the simulator takes the measured on-premise runtime $t_T$ as estimate for the execution time of $T$. It schedules $T$ on the core $c$ with the lowest $t_{max}^c$. For that core $c$, $t_{max}^c$ is updated to $t_{max}^c \leftarrow \max{(t_{max}^c, t^{ready}_T)} + t_T$.

For a task $T$ that is placed on the cloud, the simulator takes the measured runtime of the cloud UDF as an estimate for the execution time $t_T$. However, the simulator also keeps track if uplink and downlink bandwidth is occupied. The simulator assumes that each task will occupy the bandwidth fully for the amount of time required to upload/download their payloads. The earliest dispatch time $t^{dispatchable}_T$ for a cloud task is therefore determined by $t^{ready}_T$ and the earliest time when bandwidth is available. $t_{max}^{cloud}$ is updated to $t_{max}^{cloud} \leftarrow \max{(t_{max}^{cloud}, t^{dispatchable}_T)} + t_T$

The final estimate for runtime estimate $t_{max}$ is given by the maximum of $t_{max}^{cloud}$ and $t_{c}^{cloud}$ for all cores $c$.

\subsection{Simulator evaluation}
\label{app:eval-simulator}

As described in section \ref{sec: offline}, \Ours uses a simulator to estimate 

(1) the runtime of a set of tasks on a given on premise server
   
(2) the round trip time to execute tasks on the cloud.

We first evaluate these two estimations separately. We then evaluate them jointly by estimating the end-to-end runtime of a \Ours ingestion.

For the on premise estimation, we measured the runtime of a YOLO task and of a KCF task as described in section. We then estimate the runtime of the following three DAGs.

$\bullet$ YOLO: Run 60 YOLO tasks without dependencies.

$\bullet$ KCF: Run 60 KCF tasks without dependencies.

$\bullet$ Combined: Run 60 YOLO tasks, each of which feeds its output to a KCF task.  

We estimated the runtimes of these on machines with 2, 4, 8 and 16 cores. Figrue \ref{fig:simulation-res} shows the results of this experiment on the left. All estimations have an error below 9\% and the runtimes have only been overestimated. Futhermore, estimations on the same machine roughly have a similar error. Similar errors have less impact on the placement and configuration searches since all placements are overestimated similarly. We generally find that the simulation error is not a problem when looking for good placements and knob configurations. This is especially true since the selected placements are later executed to get their real runtimes for the online phase.

For the cloud estimation, we measured the round trip time of a YOLO invocation on AWS Lambda, and then invoked that function at a rate of 1 Hz for 3.5 hours. We then estimate the time at which each cloud invocation returns. Figure \ref{fig:simulation-res} shows the error of the estimation over time. 

\begin{figure}[h]
    \centering
    \includegraphics[width=0.45\textwidth]{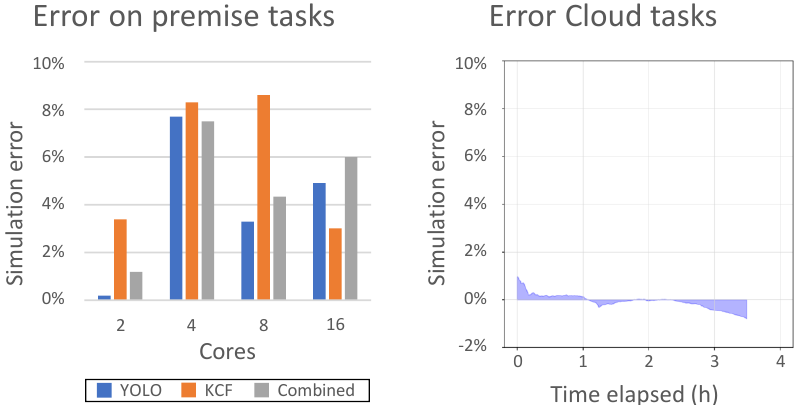}
    \vspace{-1em}
    \caption{Simulation accuracy only on on premise tasks (left) and only on cloud tasks (right)}
    \vspace{-1em}
    \label{fig:simulation-res}
\end{figure}

While there are occasional spikes for the cloud round trip times, they are so rare that they are insignificant for provisioning and therefore for the simulation. When running \Ours online, these spikes will be absorbed by the buffer which in turn causes the Knob Switcher to use more expensive placements to empty the buffer again. Since spikes occur so rarely, the additional cost is not noticeable however.

Finally, we evaluate the simulation for a run of \Ours. We hereby let the knob planner and knob switcher tune the workload's knobs and simulate the runtime of the resulting DAGs. We run \Ours on real hardware and log when each task returns. We then feed these DAGs to the simulator and let it estimate when each tasks return. Figure \ref{fig:simulation-joint} shows the estimation error over time. The simulation error was larger during rush hours as can be seen by the three spikes in the plots.

\begin{figure}[h]
    \centering
    \includegraphics[width=0.45\textwidth]{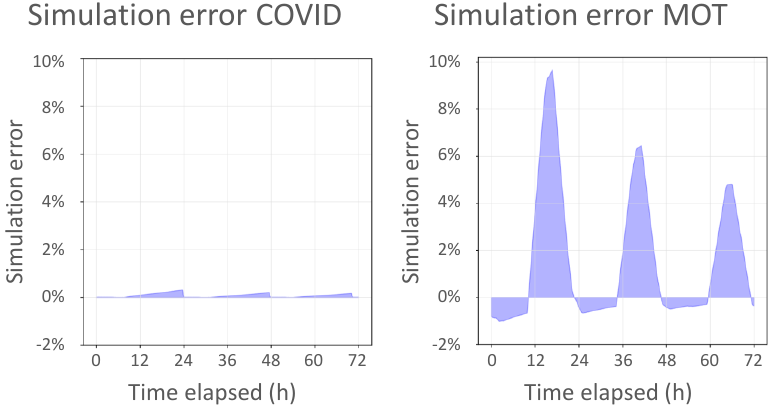}
    \vspace{-1em}
    \caption{Simulation accuracy on actual \Ours executions which combine on premise tasks and cloud tasks}
    \vspace{-1em}
    \label{fig:simulation-joint}
\end{figure}

\section{Implementation details}
\label{app_impl_details}

In the following, we motivate and give further details on the implementation of \Ours that we used in the experiments in Section~\ref{sec:eval}. This section is complementary to the implementation details provided in Section~\ref{sec:eval-impl}.

\subsection{Implementation choices}

For our experiments, we implemented \Ours on top of Ray~\cite{ray} and AWS Lambda~\cite{aws-lambda}. In the following, we briefly discuss these implementation choices.

\para{Ray} We envision \Ours's user base to mainly consist of data scientists want to use a Python API and program their UDFs in Python. The simplest way to allow this is to write \Ours in Python. However, \Ours needs to run UDFs in parallel, which is not supported natively by Python due to the Global Interpreter Lock (GIL). Therefore, we need to execute the UDFs using a third-party execution engine that is capable of executing Python functions in parallel. We chose Ray because it avoids overheads from spawning a new process for each UDF call, which is done by packages like multiprocessing~\cite{python-multiprocessing} or IPython Parallel~\cite{ipython-parallel}.

\para{AWS Lambda} We chose to execute tasks that are offloaded to the cloud on a Function-as-a-Service (FaaS) platform for simplicity. Like this, we don't need to worry about turning machines on or off (which incurs long wait times). We chose AWS Lambda since it is a well-established FaaS platform.

\subsection{Parallelization with Ray}

\begin{figure}
    \centering
    \includegraphics[width=0.48\textwidth]{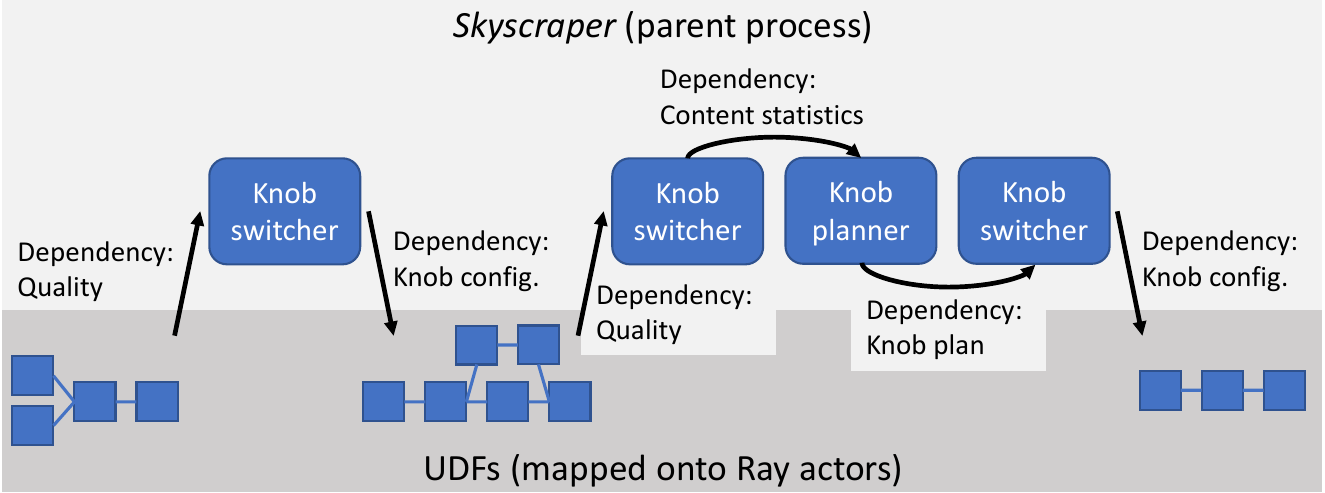}
    \caption{Visualization of dependencies in \Ours and how \Ours components are mapped onto processes.}
    \label{fig:app-dep}
\end{figure}

\Ours only maps UDFs to Ray actors and the system’s components are run inside the parent process on Python’s main thread. 

In the online phase, \Ours’s components cannot be parallelized but must instead synchronize the calls to Ray actors. Specifically, the knob switcher waits on the UDFs to finish processing the previous video segment. The knob switcher can only be run after the UDFs return because the UDFs compute their achieved quality while processing the video segment, and the knob switcher needs this quality to decide which knob configuration to use next (as described in Section~\ref{sec: online}). In our implementation, this is implemented by letting the knob switcher wait on a quality Future, whose value is set by one of the UDFs processing the previous video segment. After choosing the new knob configuration, the next UDFs are called accordingly and the knob tuner again waits on the Future to be set by one of the newly called UDFs. Similarly, the knob planner needs to wait on the statistics of the knob switcher, which record what content dynamics occurred how often. These statistics are the input to the forecasting model. The knob planner therefore waits on the future containing these statistics before it can compute a new knob plan. The knob switcher can only operate when given a knob plan, so it cannot be run in parallel to the knob planner and waits for the future containing the new knob plan. These dependencies are visualized in Figure~\ref{fig:app-dep}.

In the offline phase, by far the most time is consumed running UDFs to process the training data. Running these UDFs is parallelized by mapping them onto Ray actors (as in the online phase). Also like in the online phase, \Ours’s components are all run in the parent process because parallelizing their execution wouldn't lead to significant runtime reductions (as shown in Appendix~\ref{app:offline_scale}).

\end{document}